\documentclass[iop]{emulateapj} 

\usepackage{color}
\usepackage{float} 

\def\h2o  {H$_2$O}
\def\ho {H$_0$}

\begin{document} 
 
 \title{A More Efficient Search for H$_{2}$O Megamaser Galaxies : The Power of the X-ray and Mid-infrared Photometry}

 \author{C. Y. Kuo\altaffilmark{1}, J. Y. Hsiang\altaffilmark{2}, H. H. Chung\altaffilmark{2}, A. Constantin\altaffilmark{3}, Y.-Y. Chang\altaffilmark{4,13}, E. da Cunha\altaffilmark{5,6, 7}, D. Pesce\altaffilmark{8,9}, W. T. Chien\altaffilmark{10}, B. Y. Chen\altaffilmark{2}, J. A. Braatz\altaffilmark{11}, Ingyin Zaw\altaffilmark{12}, S. Matsushita\altaffilmark{4}, J. C. Lin\altaffilmark{1}}
 
\affil{\altaffilmark{1} Physics Department, National Sun Yat-Sen University, No. 70, Lien-Hai Rd, Kaosiung City 80424, Taiwan, R.O.C } 
\affil{\altaffilmark{2} Institute of Astronomy, National Hsing Hua University, General Building II, NTHU, No. 101, Section 2, Kuang-Fu Road, Hsinchu 30013, Taiwan, R.O.C}   
\affil{\altaffilmark{3} Department of Physics and Astronomy, James Madison University, Harrisonburg, VA 22807, USA } 
\affil{\altaffilmark{4} Academia Sinica Institute of Astronomy and Astrophysics, P.O. Box 23-141, Taipei 10617, Taiwan, R.O.C.}
\affil{\altaffilmark{5} The Australian National University, Mt Stromlo Observatory, Cotter Rd, Weston Creek, ACT 2611, Australia}
\affil{\altaffilmark{6} International Centre for Radio Astronomy Research, University of Western Australia, 35 Stirling Hwy, Crawley, WA 6009, Australia}  
\affil{\altaffilmark{7} ARC Centre of Excellence for All Sky Astrophysics in 3 Dimensions (ASTRO 3D)}  
\affil{\altaffilmark{8} Center for Astrophysics $|$ Harvard \& Smithsonian,
60 Garden Street, Cambridge, MA 02138, USA }                            
\affil{\altaffilmark{9} Black Hole Initiative at Harvard University,
20 Garden Street, Cambridge, MA 02138, USA }      
\affil{\altaffilmark{10} Department of Physics, National Taiwan University, No.1 Sec.4 Roosevelt Road Taipei 10617,Taiwan}
\affil{\altaffilmark{11} National Radio Astronomy Observatory, 520 Edgemont Road, Charlottesville, VA 22903, USA}                                                    
\affil{\altaffilmark{12} New York University Abu Dhabi, Abu Dhabi, UAE } 
\affil{\altaffilmark{12} Department of Physics, National Chung Hsing University, 40227, Taichung, Taiwan }   
\begin{abstract}
We present an investigation of the dependence of H$_{2}$O maser detection rates and properties on the mid-IR AGN luminosity, $L_{\rm AGN}$, and the obscuring column density, $N_{\rm H}$, based on mid-IR and hard X-ray photometry. Based on spectral energy distribution fitting that allows for decomposition of the black hole accretion and star-formation components in the mid-infrared, we show that 
the megamaser (disk maser) detection rate increases sharply for galaxies with 12 $\micron$ AGN luminosity $L^{AGN}_{\rm 12 \micron}$ greater than 10$^{42}$ erg~s$^{-1}$, from $\lesssim$3\%($\lesssim$2\%) to $\sim$12\%($\sim$5\%).    
By using the ratio of the observed X-ray to mid-IR AGN luminosity as an indicator of $N_{\rm H}$, we also find that the megamaser (disk maser) detection rates are boosted to 15\%(7\%) and 20\%(9\%) for galaxies with $N_{\rm H}$ $\ge$ 10$^{23}$ cm$^{-2}$ and $N_{\rm H}$ $\ge$ 10$^{24}$ cm$^{-2}$, respectively.  Combining these column density cuts with a constraint for high  $L^{AGN}_{\rm 12 \micron}$ ($\ge$10$^{42}$ erg~s$^{-1}$) predicts further increases in the megamaser (disk maser) detection rates to 19\%(8\%) and 27\%(14\%), revealing unprecedented potential increases of the  megamaser and disk maser detection rates by a factor of $7-15$ relative to the current rates, depending on the chosen sample selection criteria.  A noteworthy aspect of these new predictions is that the completeness rates are only compromised mildly, with the rates remaining at the level of $\sim$95\%($\sim$50\%) for sources with $N_{\rm H}$ $\ge$ 10$^{23}$ cm$^{-2}$ ($N_{\rm H}$ $\ge$ 10$^{24}$ cm$^{-2}$). Applying these selection methods to the current X-ray AGN surveys predicts the detection of $\gtrsim$15 new megamaser disks.

\end{abstract} 
 

\keywords{Galaxies: active -- Galaxies: nuclei -- masers -- Galaxies: surveys -- Infrared: galaxies}
 
\section{Introduction} \label{intro}
H$_{2}$O megamaser emission at $\nu \sim 22$ GHz ($\lambda \sim 1.3$ cm) originating from galactic nuclei at $\sim 0.1 - 1$ pc from the central supermassive black holes (SMBHs) provides, to date, the only known tracer of sub-parsec structures resolved in both position and velocity that appear to be associated with black hole accretion processes and, consequently, the active galactic nuclei (AGN) phenomena (e.g., Lo 2005).   
As work on the prototypical maser galaxy NGC 4258 has demonstrated (Herrnstein et al. 1999), the masing gas in such a system often resides in a subparsec scale thin disk, with the gas kinematics following nearly perfect Keplerian rotation.  As a result, one can relatively easily and accurately measure the mass of a SMBH in a H$_{2}$O maser disk (e.g. Kuo et al. 2011; Gao et al. 2017; Zhao et al. 2018).  Accurate SMBH mass measurements play a crucial role in understanding galaxy formation and evolution processes via the $M_{\rm BH}$$-$$\sigma_{*}$ relation (Ferrarese \& Merritt 2000; Gebhardt et al. 2000; G\"{u}tekin et al. 2009, Greene et al. 2016, and references therein).

Moreover, modeling a maser disk in three dimensions (e.g., Herrnstein et al. 1999; Reid et al. 2013) provides an accurate determination of the physical radius of the systemic maser component (i.e., the maser spectral component with velocities closest to the systemic velocity of the host galaxy), which can be used as a standard ruler for measuring accurate angular-diameter distances to galaxies beyond the local group.  In turn, these distance measurements enable a precise determination of the Hubble constant $H_{0}$ without inferences about the geometry of the universe (Reid et al. 2013; Kuo et al. 2013, 2015; Gao et al. 2016), thus advancing our understanding of the nature of dark energy when these measurements are used in conjunction with cosmic background radiation constraints (Hu 2005; Olling 2007).

Unfortunately, the chances of finding these golden standards are abysmally low. Of $\gtrsim 6000$ galaxy nuclei surveyed so far for 22 GHz H$_{2}$O maser emission, only 180 are detected, with $\sim$30\% of those possibly originating in disks. The typical detection rate achieved in the most extensive 22 GHz H$_{2}$O maser survey to date, the Megamaser Cosmology Project (MCP; Reid et al. 2009; Braatz et al. 2010) remains $\la 3$\% for any level of water maser emission, and $\la 1$\% for emission exhibiting a disk-like configuration. 

To enhance the detection rate of H$_{2}$O megamasers, 
systematic studies of the properties of the maser galaxies, and comparisons with their non-maser analogs have been conducted at optical wavelengths (Zhu et al. 2011; Constantin 2012, van den Bosch et al. 2016), in 2-10 keV (Greenhill et al. 2008, Zhang et al. 2006, Zhang et al. 2010; Castangia et al. 2019), and radio continuum (Zhang et al. 2012, Liu et al. 2017).  The results of these studies reveal differences between masers and non-masers, albeit with great scatter, and with new insights towards designing more efficient targeting of maser galaxies, however, without clear predictions for a significant increase in the maser detection rate (see Kuo et al. 2018 for a thorough review of these studies). 

In Kuo et al. (2018) we followed up on these works with a novel investigation of the likelihood of increasing the detection rates of H$_{2}$O megamaser emission by employing in concert the optical and mid-infrared photometric properties of the galaxies searched by the MCP with the Green Bank Telescope (hereafter the GBT galaxy sample).
We found that galaxies with water megamaser emission tend to be associated with strong emission in all of the mid-infrared bands employed by the {\it Wide-field Infrared Survey Explorer} (\emph{WISE};  Wright et al. 2010), as well as with previously proposed and newly found indicators of AGN strength in the mid-infrared, such as red $W1-W2$ and $W1-W4$ colors of the host galaxy.  The correlation with significantly red mid-IR colors also suggests that maser galaxies tend to reside in heavily obscured AGN, consistent with previous findings  (e.g. Greenhill et al. 2008; Zhang et al. 2006; Zhang et al. 2010; Castangia et al. 2013; Wagner 2013; Masini et al. 2016). Importantly, these trends predict a potential increase in megamaser detection rates to $6 - 15$\%,  depending on the specific sample selection criteria.
 
While the detection rates predicted by these new survey criteria could be boosted by a factor of $2-8$ relative to the current rates, choosing the criteria (e.g. $W1-W2 > 0.5$ \& $W1-W4>7$) which give rise to the highest detection rates (e.g. $\sim$18\%) often significantly compromises the completeness rate of H$_{2}$O maser detection (e.g. 32\%). Here, the completeness rate refers to the ratio $N^{\rm red}_{\rm maser}$/$N^{\rm tot}_{\rm maser}$, where $N^{\rm tot}_{\rm maser}$ is the total number of maser detections in the entire GBT galaxy sample and $N^{\rm red}_{\rm maser}$ indicates the number of detected maser sources showing mid-IR colors which are redder than certain thresholds. A low completeness rate suggests that a only a correspondingly small fraction of the megamasers in the GBT galaxy sample is recovered when applying certain mid-IR color cuts for galaxy selection.



Notably, four of the best-studied disk megamaser systems (i.e. UGC 3789, NGC 5765b, NGC 6264 and NGC 6323) -- all of which have well-ordered Keplerian disks suitable for an accurate $H_{0}$ determination (i.e. Reid et al. 2013; Gao et al. 2016; Kuo et al. 2013, 2015) -- exhibit blue mid-IR colors that are essentially dominated by emission from their hosts' starlight. 
While the nuclear emission of these systems shows evidence for Compton-thick absorption (i.e. the absorbing column density N$_{\rm H}$ $\gtrsim$10$^{24}$ cm$^{-2}$; Greenhill et al. 2008; Castangia et al. 2013; Masini et al. 2016), suggesting that their nuclear mid-IR colors may be intrinsically red, their AGN may not be luminous enough to dominate over the mid-IR emission from their hosts, making them ``{\it WISE} blue.'' 

Furthermore, galaxies with red mid-IR colors that correspond to the high maser detection rates we have found in Juo et al. (2018) are only a minority ($\sim$0.3\%) of the entire galaxy population, making it challenging to collect a sizable galaxy sample with these properties to start with.  As a consequence, the resulting number of possible H$_{2}$O megamaser detections from future maser surveys using such selection criteria would be small: among the $\sim$50000 galaxies iwth properties cataloged in the 2df (Colless et al. 2001), 6df (Jones et al. 2009), 2MRS (Huchra et al. 2012), RC3 (de Vaucouleurs et al. 1991), and Galaxy Zoo (Lintott et al. 2008) galaxy samples that have not yet been surveyed by the MCP, only 171 galaxies (0.3\%) have $W1-W2 > 0.5$ \& $W1-W4>7$, which would in turn yield fewer than 10 expected new disk maser detections.

Given these shortcomings, it follows that aiming at enhancing the detection rate alone is insufficient if one seeks to discover a large number of H$_{2}$O disk megamasers; to discover the bulk of the megamaser galaxy population, one must also enhance the completeness rate.  In the context of using mid-IR emission as a tool to identify megamaser candidates, the key to reaching this goal is to minimize the host galaxy contamination in the mid-IR emission in order to identify intrinsically mid-IR red (dusty) and luminous AGN.


In this paper, we explore systematic approaches that aim to boost both the maser detection rate and the completeness rate simultaneously by employing the $L^{AGN}_{12 \micron}$$-$$L^{obs}_{2-10}$ diagram, where $L^{AGN}_{12 \micron}$ and $L^{obs}_{2-10}$ refer to the 12 $\micron$ AGN luminosity (separated from that of the host via spectral energy distribution decomposition), and the observed 2-10 keV X-ray luminosity, respectively.  
We show that the use of this particular method provides an effective way to boost the maser detection rates because it allows one to select intrinsically mid-IR luminous AGN for maser search. In addition, since the ratio of $L^{obs}_{2-10}$ and $L^{AGN}_{12 \micron}$ has been shown to be an indicator of the absorbing column density of AGN (e.g. Satyapal et al. 2017), the $L^{AGN}_{12 \micron}$$-$$L^{obs}_{2-10}$ diagram provides a potential tool for selecting X-ray obscured galaxies, including Compton-thick AGN candidates which can be difficult to recognize in X-ray observations probing $\le$10 keV energy bands (e.g. Bassani et al. 1999; Cappi et al. 2006; Panessa et al. 2006; LaMassa et al. 2011; Goulding et al. 2011; Koulouridis et al. 2016). This helps to increase the completeness rate of maser survey based on mid-IR galaxy selection because maser galaxies tend to reside in heavily obscured AGN, which may not be ``{\it WISE} red".




This paper is organized as follows: 
Section 2 presents the galaxy sample and explains the methods used in compiling the X-ray and mid-IR AGN luminosities.  Section 3 explores the dependence of the maser detection rates on the 12 $\micron$ AGN luminosity and the X-ray obscuring column density, based on the $L^{AGN}_{12 \micron}$$-$$L^{obs}_{2-10}$ diagram.  The discussions and conclusions of this study are presented and discussed in Sections 4 and 5, respectively.  

Throughout this paper we adopt a $\Lambda$CDM cosmology, where $\Lambda$ denotes the cosmological constant that accounts for dark energy, and a universe that contains cold dark matter (CDM).   We adopt $\Omega_m = 0.3089$, $\Omega_{\Lambda}= 0.6911$, and \ho $= 70$ km s$^{-1}$ Mpc$^{-1}$ for the cosmological parameters.

\section{Galaxy Sample and Photometric Data Collection}

\begin{figure*}[ht]
\begin{center} 
\vspace*{0 cm} 
\hspace*{-1 cm} 
\includegraphics[angle=0, scale=0.6]{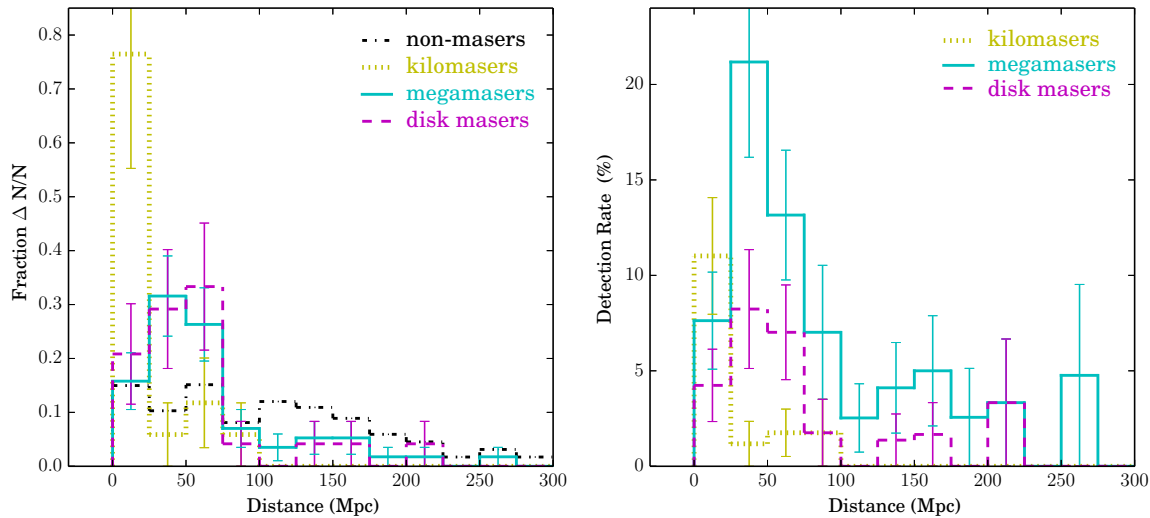}
\vspace*{0.0 cm} 
\caption{{\bf Left panel:}  Normalized number of masers and non-masers in the X-ray sample as a function of distance. {\bf Right panel:} The maser detection rates as a function of distance for sources in the X-ray sample.  Within a single bin, the detection rate for each class of maser (i.e. kilomasers, megamasers, and disk masers) is defined to be the number of masers of that class divided by the total number of X-ray sample galaxies in that bin. The error bars correspond to the associated Poisson uncertainties.}
\label{mstat}
\end{center} 
\end{figure*}

\begin{deluxetable*}{lcccccc} 
\tablewidth{0 pt} 
\tablecaption{GBT sources in the X-ray catalogs}
\tablehead{ 
\colhead{Catalog} & \colhead{Galaxy} & \colhead{Fraction} &  \colhead{$f_{\rm maser}$}   &  \colhead{$f_{\rm Mmaser}$} &  \colhead{$f_{\rm disk}$} &  \colhead{$f_{\rm nmaser}$}   \\
\colhead{Name} & \colhead{Count} & \colhead{(\%)} & \colhead{(\%)} & \colhead{(\%)} & \colhead{(\%)} & \colhead{(\%)} 
}     
\startdata 
     {\it Swift}/BAT  &              307  &        44.3 & 10.9 & 9.1 & 3.5 &  89.1 \\
     {\it XMM}-Newton  &        337  &       48.8  & 9.6 & 6.7 & 3.5 & 90.4  \\
     {\it INTEGRAL}/IBIS &        4   &       0.6 & 0.0 & 0.0 & 0.0 & 100.0  \\
     {\it Chandra} Survey &     27   &        3.9  & 11.5 & 11.5 & 0.0 &  88.5 \\
     {\it NED} catalog            &  16    &         2.3 & 37.5 & 31.2 & 18.8 &  62.5
     \enddata 
\tablecomments{ Col(1): Name of the X-ray AGN catalog; Col(2): Number of GBT sources reliably matched with X-ray sources in the catalog; Col(3): Fraction of sources relative to the entire sample of the GBT X-ray AGN; Col(4-7): Fraction of all types of masers, megamasers, disk-masers, and non-masers, respectively, relative to the whole GBT -- X-ray sample of each catalog.  } 
\end{deluxetable*}

\begin{deluxetable*}{lrrrrccccc} 
\setlength\tabcolsep{2pt} 
\tablewidth{0 pt} 
\tablecaption{Basic Properties of the GBT X-ray Sample}
\tablehead{ 
\colhead{Galaxy} & \colhead{RA} & \colhead{DEC} & \colhead{Redshift} & \colhead{Distance} & \colhead{Maser} & \colhead{log~$L^{AGN}_{\rm 12\micron}$} & \colhead{$\chi^{2}_{\nu}$} & \colhead{log~$L_{\rm OIII}$ } & \colhead{Ref.} \\
\colhead{Name} & \colhead{(degree)} & \colhead{(degree)} & \colhead{ } & \colhead{(Mpc)} & \colhead{Type} & \colhead{ (erg~s$^{-1}$) } & \colhead{ } & \colhead{ (erg~s$^{-1}$) } & \colhead{ } 
}     
\startdata 
                  UGC12914  &      0.40967  &     23.48364  &      0.01457  &     63.14  &         0  &   41.75  &   2.253  &       ---  &     ---  \\
                  UGC12915  &      0.42471  &     23.49592  &      0.01445  &     62.61  &         0  &   41.94  &   0.046  &       ---  &     ---  \\
                   NGC7811  &      0.61029  &      3.35189  &      0.02550  &    111.41  &         0  &   42.93  &   1.162  &       ---  &     ---  \\
                   Mrk1501  &      2.62919  &     10.97500  &      0.08928  &    408.03  &         0  &   44.31  &   4.487  &       ---  &     ---  \\
                     NGC17  &      2.77729  &  $-$12.10725  &      0.01978  &     85.26  &         1  &   43.17  &   0.015  &     42.34  &       5  \\
               ESO540-G001  &      8.55759  &  $-$21.43917  &      0.02683  &    117.33  &         0  &   42.97  &   0.524  &     40.84  &       5  \\
              ESO350-IG038  &      9.21958  &  $-$33.55469  &      0.02060  &     89.58  &         0  &   43.47  &   0.092  &       ---  &     ---  \\
                    NGC185  &      9.74154  &     48.33742  &   $-$0.00067  &      0.60  &         0  &   39.25  &   0.082  &       ---  &     ---  \\
                    NGC192  &      9.80598  &      0.86419  &      0.01378  &     59.68  &         0  &   41.62  &   0.189  &     40.13  &       1  \\
                    NGC214  &     10.36679  &     25.49947  &      0.01513  &     65.55  &         0  &   42.23  &   0.152  &       ---  &     ---  \\
             MCG-02-02-084  &     10.42920  &   $-$9.43953  &      0.05629  &    251.46  &         0  &   38.79  &   3.064  &       ---  &     ---  \\
                MESSIER031  &     10.68479  &     41.26903  &   $-$0.00100  &      0.69  &         0  &   37.75  &   5.986  &       ---  &     ---  \\
                   NGC232E  &     10.72004  &  $-$23.54111  &      0.02221  &     96.80  &         2  &   43.44  &   0.973  &       ---  &     ---  \\
                    NGC253  &     11.88800  &  $-$25.28836  &      0.00080  &      3.39  &         1  &   42.35  &   0.418  &       ---  &     ---  \\
                    NGC262  &     12.19642  &     31.95694  &      0.01502  &     65.11  &         2  &   43.61  &    2.57  &     41.95  &       3  \\
                    UGC524  &     12.89588  &     29.40139  &      0.03593  &    158.18  &         0  &   43.43  &   0.727  &       ---  &     ---  \\
                    NGC291  &     13.37467  &   $-$8.76778  &      0.01902  &     82.70  &         2  &   42.25  &   0.391  &     41.36  &       1  \\
                    UGC545  &     13.39558  &     12.69336  &      0.05886  &    263.42  &         0  &   44.96  &    1.61  &     41.50  &       5  \\
                    NGC315  &     14.45368  &     30.35250  &      0.01647  &     71.47  &         0  &   42.41  &   2.674  &       ---  &     ---  \\
                    Mrk352  &     14.97200  &     31.82692  &      0.01485  &     64.37  &         0  &   42.12  &   1.545  &     38.97  &       4  \\
                      UM85  &     16.68859  &      6.63389  &      0.04097  &    181.03  &         0  &   43.39  &   1.407  &       ---  &     ---  \\                     
                   
\enddata 
\tablecomments{ Col(1): name of the galaxy; Col(2) \& Col(3): the galaxy coordinates (in degrees, J2000); Col(4): redshift; Col(5): luminosity distance in Mpc; Col(6); the type of the maser system. Here, the number 0, 1, 2, and 3 refer to non-maser, kilomaser, non-disk megamaser, and disk megamaser, respectively; Col(7):  the best-fit 12$\micron$ AGN luminosity from the SED fitting; Col(8):  the reduced $\chi^{2}$ of the fit; Col(9):  The [OIII] $\lambda$5007 luminosity $L_{\rm [OIII]}$ (erg~s$^{-1}$). In all sources with $L_{\rm [OIII]}$ available from literature, we applied the internal reddening corrections using the Balmer decrement, assuming an intrinsic ratio of (H$\alpha$/H$\beta$)$_{0}$ of 3.0 and a H$\beta$/H$\alpha$ color index for extinction of 2.94 (Bassani et al. 1999); Col(10): Reference for the $L_{\rm [OIII]}$ measurement : 1. The MPA-JHU DR8 release of spectrum measurements\tablenotemark{a}; 2. Ho et al. (1997); 3. Shu et al. (2007); 4. Ueda et al. (2015); 5. Malkan et al. (2017); 6. Moustakas et al. (2006); 7. Parisi et al. (2014); 8. Zhu et al. (2011);  9. Dahari et al. (1988); 10. Masetti et al. (2008); 11. Robitaille et al. (2007); 12. Panessa et al. (2006). \newline
Only a portion of the table is shown here to demonstrate its form and content. A machine-readable version of the full table is available. } 
\tablenotetext{a}{https://www.sdss3.org/dr10/spectro/galaxy\_mpajhu.php}
\end{deluxetable*}

\begin{deluxetable*}{lcclllcc} 
\setlength\tabcolsep{3pt} 
\tablewidth{0 pt} 
\tablecaption{The X-ray Properties of the GBT X-ray Sample}
\tablehead{ 
\colhead{Galaxy} & \colhead{$L_{2-10}^{obs}$} & \colhead{$L_{2-10}^{int}$} & \colhead{log~$N_{\rm H}$} & \colhead{$\Gamma_{1}$} & \colhead{$\Gamma_{2}$} & \colhead{$\chi^{2}_{\nu}$} & \colhead{Ref.} \\
\colhead{Name} & \colhead{(erg~s$^{-1}$) } & \colhead{(erg~s$^{-1}$)  } & \colhead{(cm$^{-2}$) } & \colhead{ } & \colhead{ }  & \colhead{ } & \colhead{ } 
}     
\startdata 
                  UGC12914  &        40.29  &        40.29  &                20.11  & 1.68$^{+0.36}_{-0.28}$  &       ---  &     0.973  &       2  \\
                  UGC12915  &        40.24  &        40.24  &                20.14  & 1.78$^{+0.48}_{-0.23}$  &       ---  &     1.026  &       2  \\
                   NGC7811  &        42.85  &        42.85  & 20.00$^{+0.00}_{-0.00}$  & 2.23$^{+0.06}_{-0.06}$  &       ---  &       ---  &       1  \\
                   Mrk1501  &        44.19  &        44.19  & 21.04$^{+0.07}_{-0.04}$  & 1.73$^{+0.03}_{-0.07}$  &       ---  &       ---  &       1  \\
                     NGC17  &        41.43  &        41.44  &                20.98  & 0.51$^{+0.16}_{-0.20}$  &       ---  &     1.172  &       2  \\
               ESO540-G001  &        41.74  &        41.74  &                20.00  & 1.97$^{+0.05}_{-0.05}$  &       ---  &     1.160  &       2  \\
              ESO350-IG038  &        40.95  &        40.95  &                20.37  & 2.17$^{+0.17}_{-0.20}$  &       ---  &     1.814  &       2  \\
                    NGC185  &     $<$34.97  &     $<$34.97  & 20.98$^{+0.45}_{-0.07}$  &                  ---  &       ---  &     1.272  &       2  \\
                    NGC192  &        40.29  &        40.29  & 21.41$^{+0.07}_{-0.09}$  & 1.44$^{+0.28}_{-0.24}$  &       ---  &     1.193  &       2  \\
                    NGC214  &        40.80  &        40.80  &                20.11  & 0.98$^{+0.23}_{-0.21}$  &       ---  &     2.341  &       2  \\
             MCG-02-02-084  &     $<$41.86  &     $<$41.86  & 20.00$^{+1.00}_{-0.00}$  &                  ---  &       ---  &     1.200  &       2  \\
                MESSIER031  &        37.60  &        37.60  & 20.00$^{+0.57}_{-0.90}$  & 1.89$^{+0.17}_{-0.16}$  &       ---  &     1.202  &       2  \\
                   NGC232E  &        42.65  &        43.22  & 23.50$^{+0.07}_{-0.07}$  & 1.76$^{+0.11}_{-0.42}$  &       ---  &       ---  &       1  \\
                    NGC253  &        39.50  &        39.51  & 21.30$^{+0.02}_{-0.02}$  & 2.13$^{+0.03}_{-0.03}$  &       ---  &     1.886  &       2  \\
                    NGC262  &        43.14  &        43.44  & 23.12$^{+0.03}_{-0.02}$  & 1.51$^{+0.12}_{-0.12}$  &       ---  &       ---  &       1  \\
                    UGC524  &        42.99  &        42.99  & 20.00$^{+0.00}_{-0.00}$  & 1.68$^{+0.11}_{-0.11}$  &       ---  &       ---  &       1  \\
                    NGC291  &        40.83  &        40.83  &                20.00  & 1.33$^{+0.21}_{-0.36}$  &       ---  &     1.145  &       2  \\
                    UGC545  &        43.64  &        43.63  & 20.00$^{+0.06}_{-0.17}$  & 2.02$^{+0.01}_{-0.01}$  &       ---  &     1.260  &       2  \\
                    NGC315  &        41.63  &        41.67  & 21.71$^{+0.14}_{-0.20}$  & 2.03$^{+0.14}_{-0.13}$  &       ---  &     1.354  &       2  \\
                    Mrk352  &        42.74  &        42.74  & 20.00$^{+0.00}_{-0.00}$  & 1.92$^{+0.03}_{-0.02}$  &       ---  &       ---  &       1  \\
                      UM85  &        42.82  &        43.41  & 23.81$^{+0.22}_{-0.25}$  & 2.22$^{+0.13}_{-0.76}$  &       ---  &       ---  &       1  \\
                   NGC6251  &        42.53  &        42.75  & 23.29$^{+0.14}_{-0.08}$  & 1.67$^{+0.11}_{-0.15}$  & 3.31$^{+0.63}_{-0.62}$  &     1.228  &       2  \\
                   NGC6217  &        39.53  &        39.53  &                20.18  & 2.72$^{+0.19}_{-0.19}$  &       ---  &     1.422  &       2  \\
    2MASXJ16383091-2055246  &        42.92  &        42.74  & 23.58$^{+0.09}_{-0.06}$  & 2.09$^{+0.79}_{-0.60}$  & 2.18$^{+0.05}_{-0.05}$  &     1.282  &       2  \\   
            UGC10814NOTES01  &        42.81  &        42.82  & 21.32$^{+0.13}_{-0.17}$  & 1.41$^{+0.08}_{-0.08}$  &       ---  &       ---  &       1  \\
                     3C353  &        42.32  &        42.46  & 22.85$^{+0.08}_{-0.07}$  & 1.85$^{+0.34}_{-0.25}$  & 0.81$^{+0.47}_{-0.34}$  &     1.266  &       2  \\   
\enddata 
\tablecomments{ Col(1): name of the galaxy; Col(2): The observed AGN luminosity in the 2$-$10 keV energy band; Col(3): the absorption-corrected intrinsic AGN luminosity in the 2$-$10 keV energy band. Except for sources from the {\it Swift}/BAT catalog, Shu et al. (2007) and LaMassa et al. (2011), the absorption correction was calculated purely based on the best-fit $N_{\rm H}$ value without considering whether the source is Compton-thick. Therefore, $L_{2-10}^{int}$ for Compton-thick sources in non-{\it Swift}/BAT catalogs may be underestimated; Col(4): Neutral hydrogen absorption column density $N_{\rm H}$ (cm$^{-2}$).  For sources from the {\it XMM-Newton} catalog, the $N_{\rm H}$ value without uncertainty provided indicates that the column density was fixed to the value shown in the table in the spectral fitting. For sources from non-{\it XMM-Newton} catalogs, the lack of uncertainty means that the measurement error is not available from the corresponding reference; Col(5): The photon index $\Gamma$ obtained from the X-ray spectral-fitting; Col(6): The photon index for the second power-law fit to the data. This column have values only for {\it XMM-Newton} sources which are best-fit by the double powerlaw model; Col(7):  the reduced $\chi^{2}$ of the X-ray spectral-fitting for sources from the {\it XMM-Newton} catalog; Col(8): reference for the X-ray properties :  1. Ricci et al. (2017); 2. Coral et al. (2015); 3. She, Ho, \& Feng (2017); 4. Malizia et al. (2012); 5. Malizia et al. (2016); 6. Akylas et al. (2009); 7. Brightman et al. (2011); 8. Chitnis et al. (2009); 9. Gliozzi et al. (2008); 10. Guainazzi et al. (2005); 11. Gonz$\acute{\rm a}$lez-Mart$\acute{\rm i}$n et al. (2006); 12. Noguchi et al. (2009); 13. Turner et al. (1989); 14. LaMassa et al. (2011); 15. She et al. (2007); 16. Castangia et al. (2013). \newline
Only a portion of the table is shown here to demonstrate its form and content. A machine-readable version of the full table is available.  } 
\end{deluxetable*}

\subsection{Galaxy Sample and the X-ray Data}
The galaxies we study in this paper are drawn from the GBT sample studied by Kuo et al. (2018), which represents the largest and most comprehensive catalog\footnote{The catalog can be accessed via the weblink: https://safe.nrao.edu/wiki/bin/view/Main/MegamaserProjectSurvey} of galaxies
surveyed for water maser emission at 22 GHz. This sample contains 4836 galaxies surveyed by the GBT prior to September 2016. These galaxies are primarily narrow-emission line AGN selected from galaxy catalogs including the Sloan Digital Sky Survey (SDSS; York et al. 2000; Abazajian et al. 2009), 2MASS Redshift Survey (Huchra et al. 2012), 2dF Survey (Colless et al. 2001), and 6dF Survey (Jones et al. 2009), via their optical emission line ratios using the BPT diagram method (Baldwin, Phillips \& Terlevich, 1981), with no additional criteria such as galaxy colors or magnitudes. 


The redshifts of the GBT sample galaxies range from 0.0 to 0.07, and the overall maser detection rate is $\sim$3\%. Hereafter, we refer to galaxies with confirmed H$_{2}$O maser emission as {\it maser galaxies} or {\it masers}\footnote{Of all the 180 megamaser detections mentioned in the introduction, two are found associated with the high redshift quasars MG J0414$+$0534 (z$=$2.64; Impellizzeri et al. 2008) and SDSS J080430.99$+$360718.1 (z$=$ 0.66; Barvainis \& Antonucci 2005). These two maser sources are not discovered as part of the GBT maser survey and we do not included them in the analysis presented in this paper.}, and we refer to galaxies with no maser detection as {\it non-maser galaxies} or {\it non-masers}.  Within the category of masers, we further define three subsamples called {\it kilomasers}, {\it megamasers}, and {\it disk masers}. Kilomasers and megamasers refer to masers having an isotropic H$_{2}$O luminosity $L_{\rm H_{2}O}$ $<10$  $L_{\odot}$ and $L_{\rm H_{2}O}$ $>10$  $L_{\odot}$, respectively. Disk masers are megamaser systems whose maser spectra display characteristic spectral structures that suggest the presence of subparsec scale maser disks; thus, the {\it megamasers indicate a maser population which includes both disk masers and non-disk megamasers.} (see Section 2 in Kuo et al. 2018 for a more detailed description of these types of masers). The isotropic H$_{2}$O luminosities for all of the maser galaxies can be found in Table 1 of Kuo et al. (2018).


To obtain literature values for the observed 2$-$10 keV X-ray luminosities $L_{2-10}^{obs}$ for galaxies in the GBT sample, we prioritize those X-ray surveys which provide X-ray properties based on X-ray spectral-fitting.  The X-ray spectral analysis is important for our work because our investigation primarily relies on the {\it intrinsic} X-ray luminosities of AGN whose non-thermal continuum emission can be described by power-law distributions in their X-ray spectra.  Spectral decomposition is necessary to separate the power-law component from the thermal emission (e.g. soft excess). When the signal-to-noise of an  X-ray spectrum is high enough (e.g. spectra from the {\it Swift}/BAT survey described below), we adopt data in which any existing Fe K-alpha line components are removed.  


After assembling a set of X-ray surveys which provide 2-10 keV X-ray luminosities based on X-ray spectral analysis, we cross-matched the GBT sample with the four largest X-ray source catalogs in our collection. These X-ray catalogs include (1) the {\it Swift}/BAT 70-month AGN catalog (Ricci et al. 2017), (2) the {\it XMM-Newton} spectral-fit database (Corral et al. 2015), (3) the {\it Chandra} Survey of Nearby Galaxies (She, Ho, \& Hua 2017), and (4) the {\it INTEGRAL}/IBIS AGN catalogue (Malizia et al. 2012, 2016). We also searched the 40-month catalog of the NuSTAR Serendipitous Survey (Lansbury et al. 2017) for counterparts of the GBT galaxy sample, but we didn't find any matches. 

In addition to the above large X-ray source catalogs, we also cross-matched the GBT sample with smaller X-ray catalogs reported in NASA/IPAC Extragalactic Database (NED).  
Upon extensively searching for all references of X-ray photometric data for each galaxy in the GBT sample, we identified a set of measurements that we collectively refer to here as the {\it NED} catalog.  The number of GBT galaxies found in each X-ray catalog, including a split per maser type, are listed in Table 1.

In cross-matching sources between the X-ray catalogs we use a matching radius of 10 arcseconds, from which we identify 691 X-ray counterparts of the GBT sample. To ensure reliable cross-matching and compatibility with additional optical and near-IR photometric data used in spectral energy distribution (SED) fitting, we further  require the matched X-ray counterparts to be within the effective (i.e., half-light) radii $r_{\rm eff}$ of the GBT galaxies, with $r_{\rm eff}$ obtained from the SDSS or the 2MASS galaxy surveys (see Section 2.2.2 for details).  This multi-wavelength, multi-catalog search leads to successful cross-matching of 
642 galaxies ( hereafter the {\it X-ray sample}) that have all the necessary broad-band rest-frame UV-to-mid-IR photometry for measuring an intrinsic mid-IR AGN luminosity based on SED decomposition (see Section 2.2). The majority (93\%) of the X-ray sample sources are included in the {\it Swift}/BAT catalog (44\%) and the {\it XMM-Newton} spectral-fit database (49\%).   The total numbers of kilomasers, megamasers, and disk masers in the X-ray sample are 15 (2.3\%), 53 (8.3\%), and 24 (3.7\%), respectively.

Table 2 lists the coordinates, redshifts, distances, and the mid-IR (at $12 \mu$m) and optical (in [OIII] emission) luminosities of the X-ray sample sources. The X-ray properties of these sources, which include the 2-10 keV luminosities, neutral hydrogen absorbing column densities $N_{\rm H}$, and the photon index values $\Gamma$ (from fitting power-laws to the X-ray spectra) are shown in Table 3. The distance distribution of these X-ray sources and the maser detection rates as a function of distance are shown in the left and right panels of Figure \ref{mstat}, respectively. In Section 4.2, we explore why the detection rate of megamasers in the X-ray sample has a different distance distribution with respect to the GBT sample. The following subsections present in detail the considerations for retrieving the necessary data from the above-listed X-ray catalogs.


\subsubsection{Swift/BAT 70-month AGN catalog}
The {\it Swift}/BAT source catalog consists of 838 AGN detected in the 70-month {\it Swift}/BAT all-sky survey, which probes X-ray sources in the 14$-$195 keV energy range (Ricci et al. 2017). The spectral analysis used in this catalog combines X-ray data from {\it XMM-Newton}, {\it Swift}/XRT, {\it ASCA}, {\it Chandra}, and {\it Suzaku}
observations  in the X-ray band covering 0.3$-$10 keV.  Because of the broadband nature (0.3$-$150 keV) of the analysis, the {\it Swift}/BAT AGN catalog provides the most accurate measurements of the X-ray properties for GBT sample sources included in this catalog. In particular, since {\it Swift}/BAT can detect X-ray photons with energies beyond 10 keV (which can traverse heavy obscuration), measurements of $N_{\rm H}$ from this catalog are far less susceptible to systematic biases caused by large absorbing columns in Compton-thick AGN than observations that only probe the $\leq$10 keV energy bands (e.g. {\it Chandra} \& {\it XMM-Newton}).

The X-ray spectral analysis in the {\it Swift}/BAT catalog was carried out for two categories of sources (i.e. nonblazar AGN and blazars) using a series of models of successive complexity. For nonblazar AGN, which are further subdivided into obscured and unobscured sources, 17 different spectral decomposition models were fit to the 0.3$-$150 keV broadband data.  For blazars, 7 spectral models were adopted to fit the data.  For all of the {\it Swift}/BAT sources, the broadband data enables the X-ray spectral analysis to remove the Fe K$\alpha$ component and various thermal components in the observed X-ray luminosities $L^{\rm obs}_{2-10}$, ensuring that the AGN luminosities listed in Table 3 only include the power-law component of the AGN X-ray continuum emission. 

Because of the reliability of its X-ray parameter estimation, we give higher priority to the {\it Swift}/BAT data when a GBT sample galaxy appears in multiple X-ray catalogs including the Swift/BAT. Among the 4836 galaxies in the GBT sample, 307 have reliable matches to the Swift/BAT 70-month catalog. For a few sources in our sample which have X-ray counterparts in the {\it Swift}/BAT survey, instead of using $L^{\rm obs}_{2-10}$ provided by the {\it Swift}/BAT catalog, we adopt the estimated 2$-$10 keV luminosities $L^{\rm estimate}_{2-10}$ obtained by transforming the observed 14$-$195 keV luminosities from the {\it Swift}/BAT survey ($L^{\rm obs}_{14-195}$) into 2$-$10 keV luminosities using a correction factor of 0.37 (i.e. $L^{\rm estimate}_{2-10}$ $\equiv$ $0.37{\times}L^{\rm obs}_{14-195}$; see Koss et al. 2017)\footnote{The conversion factor 0.37 corresponds to an X-ray spectrum photon index $\Gamma$ of 1.8.}. As indicated in Figure 15 in Ricci et al. (2017), $L^{\rm estimate}_{2-10}$ shows a strong correlation with $L^{\rm obs}_{2-10}$ for objects with log$N_{\rm H}$ $\lesssim$23.7, with a scatter of $\sim$0.3 dex which is likely caused by differences in the shapes of the X-ray continuum spectra\footnote{The scatter in log$L^{\rm estimate}_{2-10}$ will be 0.32 if the X-ray continuum spectra of a sample of X-ray AGN are fully described by power-law distributions with the photon indices $\Gamma$ following a Gaussian distribution with a mean of 1.76 and a standard deviation of 0.29 (i.e. the mean and the standard deviation of $\Gamma$ for {\it Swift}/BAT sources in the X-ray sample.)} and by intrinsic flux variability of the X-ray emission (Ricci et al. 2017). However, a few sources have absorption-corrected 2$-$10 keV fluxes that are considerably higher than those expected from the observed 14$-$150 keV flux. According to Ricci et al. (2017), their $L^{\rm obs}_{2-10}$ might be overestimated. Given the possibility of significant overestimation, we adopt $L^{\rm estimate}_{2-10}$ as the observed 2$-$10 keV luminosities for sources in our sample with $L^{\rm obs}_{2-10}$  $>$ 5$\times$$L^{\rm estimate}_{2-10}$.

\begin{figure*}[ht]
\begin{center} 
\vspace*{0 cm} 
\hspace*{-0.3 cm} 
\includegraphics[angle=0, scale=0.63]{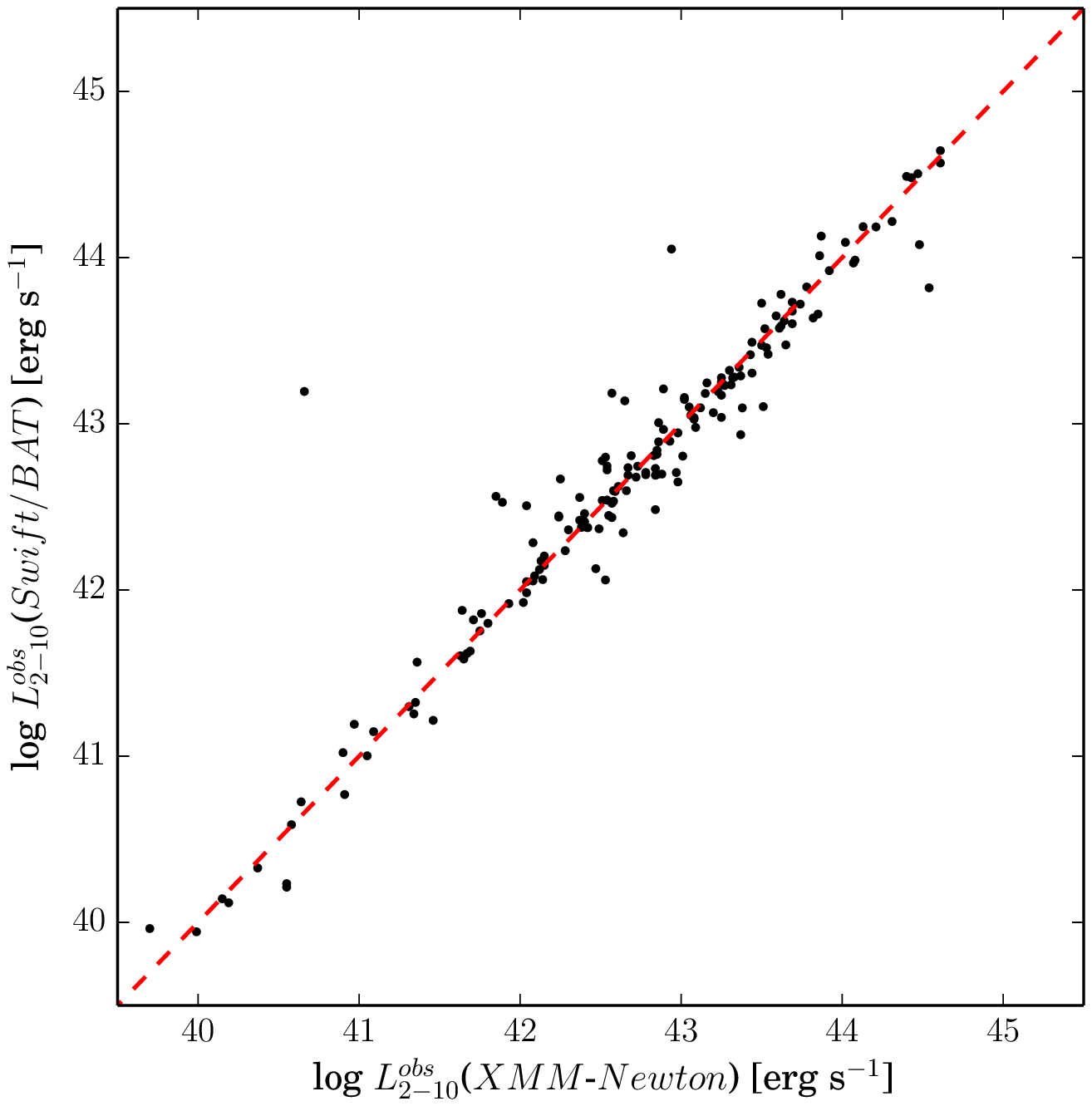}
\vspace*{0.0 cm} 
\caption{The comparison of the observed 2-10 keV X-ray luminosities of the 168 sources which exist in both the {\it XMM-Newton} spectral-fit database and the {\it Swift}/BAT AGN catalog. The horizontal and vertical axes show the measurements from the {\it XMM-Newton} and {\it Swift}/BAT catalogs, respectively. Except for a few outliers (see text), the observed X-ray luminosities measured from these two X-ray surveys are well consistent with each other, with a mean scatter of 0.16 dex.}
\label{xray_comp}
\end{center} 
\end{figure*}

\subsubsection{the {\it XMM-Newton} Spectral-fit Database}
The {\it XMM-Newton} spectral-fit database (XMMFITCAT; Corral et al. 2015) provides X-ray spectral fitting results for sources detected in the {\it XMM-Newton} serendipitous survey (Rosen et al. 2016), which probes the 0.2$-$12 keV energy band. In its most recent release (3MM-DR6), XMMFITCAT includes fitting results for 146825 detections.  Among all unique {\it XMM-Newton} detections, we found 505 matches between XMMFITCAT and the GBT sample, with 168 sources also belonging to the {\it Swift}/BAT AGN catalog. Therefore, XMMFITCAT provides X-ray information for 337 independent X-ray sources for our study.

Unlike the {\it Swift}/BAT catalog, XMMFITCAT does not directly provide a measurement of $L^{\rm obs}_{2-10}$. Instead, it provides best-fit parameters for six spectral models (Corral et al. 2015). To obtain $L^{\rm obs}_{2-10}$, we adopt the model parameters from the ``preferred model" (Corral et al. 2015) recommended by XMMFITCAT, which is selected according to the goodness of each fit, which we then use to estimate $L_{2-10}^{obs}$ from the power-law component of the X-ray emission following the equation
\begin{equation}
L_{2-10}^{obs} = 4\pi D_{\rm L}^{2}\int^{10 keV}_{2 keV} e^{-N_{\rm H}\sigma(E)} NE^{1-\Gamma} dE~,
\end{equation}
where $D_{\rm L}$ is the luminosity distance, $\sigma(E)$ is the photoelectric cross-section from Morrison \& McCammon (1983), E is the photon energy in units of keV, and $N_{\rm H}$, $N$, and $\Gamma$ are the best-fit absorbing column density, normalization factor, and photon index given by XMMFITCAT, respectively. When the double power-law model (i.e. model 5) is selected as the preferred model, we evaluate $L_{2-10}^{obs}$ by summing the luminosities of the two power-law components. When the preferred model has no power-law component and purely includes thermal emission (i.e. model 1), we use the 2-10 keV X-ray luminosity of the thermal component as an upper limit on $L^{\rm obs}_{2-10}$. There are nine sources in this last category, and they are indicated as upper limits for both $L^{obs}_{2-10}$ and $L^{int}_{2-10}$ in Table 3.

Given the significant overlap between the XMMFITCAT and {\it Swift}/BAT AGN catalogs, we are able to check for consistency in the $L^{obs}_{2-10}$ measurements.  Figure \ref{xray_comp} compares the luminosities obtained from data appearing in both catalogs, where it is readily apparent that with the exception of a few outliers, there is good general consistency between the two sources of $L^{\rm obs}_{2-10}$ measurements (the two main outliers for which $L^{obs}_{2-10}$($Swift$/BAT) are substantially higher than $L^{obs}_{2-10}$($XMM$-Newton) are 3C84 and 2MASX J05580206$-$3820043, which are non-masers). The mean scatter between the two different measurements is  0.16 dex, well within the scatter ($\sim$0.3 dex) seen in the correlation between $L^{\rm estimate}_{2-10}$ and $L^{\rm obs}_{2-10}$ (see Section 2.1.1) which is partly caused by the intrinsic flux variability (Ricci et al. 2017).

\subsubsection{ {\it INTEGRAL}/IBIS survey}
The latest version of the {\it INTEGRAL}/IBIS AGN catalog (Malizia et al. 2012, 2016) consists of 363 high-energy emitters confirmed to be AGNs. This catalog provides 2-10 keV fluxes and absorbing column densities for these sources based on spectral-fitting. We found 147 counterparts in the GBT sample, with 143 of them being either included in the {\it Swift}/BAT AGN catalog or XMMFITCAT.  Thus, cross-matching with the {\it INTEGRAL}/IBIS AGN catalog only provides X-ray information for 4 additional sources.

\subsubsection{The {\it Chandra} Survey of Nearby Galaxies}
This X-ray catalog is derived from the {\it Chandra} survey of nearby galaxies (She, Ho, \& Feng 2017), and it contains 314 AGN within 50 Mpc. The majority of these sources are low-luminosity AGN. X-ray spectra were extracted for 154 AGN having photon counts $\ge$100 in the energy band 0.3$-$8 keV. For these sources, She, Ho, \& Feng (2017) fit either a simple absorbed power-law model, a power-law model plus a thermal component, or a partial absorption model to the spectral data to determine the X-ray properties. For the 160 sources without sufficient photon counts (i.e. $<$100), the authors estimate the 2$-$10 keV flux and luminosity with their source count rate assuming an absorbed power-law model, and they infer the absorbing column $N_{\rm H}$ based on the observed hardness ratio.

Here, we consider only the 154 AGN with available spectral-fitting results. After cross-matching with the GBT sample and removing sources already included in the previous catalogs,  the {\it Chandra} AGN catalog provides 27 additional AGN for our current study. For most of these sources (24 out of 27), the X-ray spectra can be fit with either a simple absorbed power-law or with a power-law model plus a thermal component; for these sources we evaluate $L^{\rm obs}_{2-10}$ using Equation (1), based on the spectral-fitting results provided by Table 4 in She, Ho, \& Feng (2017). For the remaining (3 out of 27) sources that are best-fit by the partial absorption model, we requested the partial covering factor from Rui She (private communication).  We use this additional information to evaluate the 2$-$10 keV luminosity based on
\begin{equation}
L_{2-10}^{obs} = 4\pi D_{\rm L}^{2}\int^{10 keV}_{2 keV} [\eta e^{-N_{\rm H}\sigma(E)}+(1-\eta)] NE^{1-\Gamma} dE~,
\end{equation}
where $\eta$ is the dimensionless covering fraction (0 $<$$\eta$~$\le$1).

\subsubsection{NED Catalog}
To identify additional X-ray measurements, particularly estimates of the absorbing column density for GBT sample galaxies not included in the four large catalogs discussed above, we searched the references provided by NED.  Our extensive literature search resulted in X-ray photometric data for an additional 39 sources in the GBT sample, of which 16 galaxies have $L^{\rm obs}_{2-10}$ and $N_{\rm H}$ derived from spectral-fitting (see the references in Table 2). For the 23 sources without spectral fits, the references provide only the total 2$-$10 keV luminosity, which may contain thermal emission in this energy band. As a result, we do not use these measurements to infer $N_{\rm H}$ from the $L^{AGN}_{12 \micron}-L^{obs}_{2-10}$ diagram so as not to contaminate it with underestimates of the column density (upper limits are not necessarily relevant for this study).

\subsection{Mid-infrared AGN luminosity}
\subsubsection{The Methods}
Studies of mid-IR$-$X-ray relations in AGNs often determine mid-IR AGN luminosities ($L_{\rm MIR}$) by performing high-resolution (i.e., subarcsecond) imaging of the target sources in these wavelengths (e.g. Fiore et al. 2009; Gandhi et al. 2009; Goulding et al. 2011; Sazonov et al. 2012; Asmus et al. 2014, 2015, 2016). High-resolution imaging permits high contrast with the host galaxy light, thus ensuring that the AGN dominates the mid-IR emission in the central region of the source. This generally leads to an accurate measurement of $L_{\rm MIR}$ (or simply the 6 $\micron$ or 12 $\micron$ luminosity), which has been shown to be tightly correlated with the absorption-corrected intrinsic 2-10 keV X-ray luminosity $L_{2-10}^{int}$ of the central AGN (the $L_{\rm MIR}$$-$$L_{2-10}^{int}$ relations; Lutz et al. 2004; Fiore et al. 2009; Gandhi et al. 2009; Sazonov et al. 2012; Asmus et al. 2015). One of the most well-known relations was found by Gandhi et al. (2009; the {\it Gandhi relation} hereafter), who discovered a tight correlation between the 12 $\micron$ AGN luminosity $L^{AGN}_{\rm 12 \micron}$ $\equiv$ $\nu L_{\nu}$(12$\micron$) and the intrinsic 2-10 keV luminosity $L_{2-10}^{int}$.   

An alternative way to reliably measure the $L_{\rm MIR}$ of an AGN is to fit SED templates to broadband photometric observations of the source, which allow for a decomposition of the total observed mid-IR emission into components associated with the AGN and with the host galaxy (e.g. Georgantopoulos et al. 2011; Mateos et al. 2015; Koulouridis et al. 2016). This approach is particularly useful for galaxy samples where  high-resolution mid-IR imaging is not available, as is the case for our GBT sample.  To follow this approach, we perform SED decomposition for our sources using a recent version of the SED fitting code {\it MAGPHYS} (da Chunha et al. 2008, 2015; Chang et al. 2017) applied to broadband UV-to-mid-IR photometric data.

\subsubsection{The SED Decomposition}

In {\it MAGPHYS}, the AGN emission is reproduced from a set of empirical templates (see Figure 2 in Chang et al. 2017) from Mullaney et al. (2011; template 1 \& 2, corresponding to the low and high luminosity Seyfert 2s), Richards et al. (2006; template 3, for the empirical QSO template), and Polletta et al. (2007; template 4, for the average Seyfert 1 galaxies), which span in a representative way the global range of known AGN SEDs. To avoid degeneracies in the SED fitting, we only use four typical templates. 

The broadband photometric data used in our SED fitting includes data taken in the UV, optical, near-IR, and mid-IR wavelengths, with the UV-to-optical data obtained from Data Release 14 (DR14) of the {\it Sloan Digital Sky Survey} (SDSS) (e.g. Abolfathi et al. 2018), the near-IR data from the {\it Two Micron All Sky Survey} (2MASS; Skrutskie et al. 2006), and the mid-IR data from {\it Widefield Infrared Survey Explorer} (WISE; Wright et al. 2010). We obtain the SDSS data via the Skyserver DR14\footnote{http://cas.sdss.org/dr14}, and access the 2MASS and WISE data via NASA/IPAC Infrared Science Archive using the IRSA7 catalog tool.

The final product of the {\it MAGPHYS} SED fitting is a set of best-fit parameters, including the infrared AGN luminosity $L_{\rm IR, AGN}$ in the wavelength range 3$-$2000 $\micron$. Since we are mainly interested in the 12 $\micron$ AGN luminosity $L^{AGN}_{\rm 12 \micron}$, we measure this value by using the conversion factors $\epsilon$ $\equiv$ $L_{\rm IR, AGN}$/$L^{AGN}_{\rm 12 \micron}$ of 3.07, 2.49, 2.62, and 2.50 for templates 1, 2, 3, and 4, respectively, which we have calculated directly from the four SED templates. 

The following section provides details on the data collection and strategies for minimizing host galaxy contamination in the 12 $\micron$ AGN luminosity measurements.

\subsubsection{The Broad-band UV-to-mid-IR Photometry}

To obtain the broadband photometry, we cross-match the X-ray sample with the SDSS, 2MASS, and WISE catalogs using a matching radius of 6\arcsec (i.e. the resolution of WISE at 3.4$\micron$ and 4.6$\micron$). We use a smaller matching radius here because we assume that the position accuracies of galaxies in the optical and infrared catalogs are better than those in the X-ray catalogs. We note that increasing the matching radius to 10\arcsec does not increase the number of successful cross-matches and therefore does not affect our results. 

Among the source catalogs provided by 2MASS, we adopt the {\it extended source catalog} (XSC) for cross-matching. To minimize the chance of mismatch, we further require the X-ray sample galaxies to be within the effective radii $r_{\rm eff}$ of the matched sources, with $r_{\rm eff}$ determined primarily from the 2MASS catalog (88\% galaxies). When $r_{\rm eff}$ is not available in 2MASS XSC for a particular source (12\% of the X-ray sample), we adopt the value from the SDSS catalog for making the comparison. Based on this cross-matching procedure, we find that there are 291 galaxies (45\%) in the X-ray sample that do not have UV-optical photometry available from the SDSS catalog. For these sources, we perform the SED fitting using only the photometry from the near-IR and mid-IR bands.


Following the procedure described in Mateos et al. (2015), we adopt the SDSS model magnitudes in the $u,g,r,i,z$ (i.e., UV-to-optical) bands, while for the 2MASS JHK bands (at 1.25, 1.65, and 2.17 $\micron$, respectively) we use the fiducial Kron elliptical aperture magnitudes.  For both the SDSS and 2MASS data, we applied a Galactic extinction correction according to Schlafly \& Finkbeiner (2011).  When evaluating the global galaxy flux densities in the WISE bands (i.e., $W_1$, $W_2$, $W_3$, $W_4$, at 3.4, 4.6, 12, 22 $\micron$, respectively), we use the elliptical aperture magnitudes ($w_{\rm gmag}$) for the X-ray sample sources that are flagged as extended (constituting 99\% of the sample), and we adopt profile-fit photometric magnitudes for the remaining 1\% of sources that are spatially unresolved in all of the WISE bands.


To account for possible systematic errors in the photometric data, we adopt the systematic uncertainties of 0.1 mag used by Chang et al. (2015) for the SDSS and WISE bands, and we add 0.05 mag of systematic uncertainty in quadrature with the photometric errors for the 2MASS bands. For the WISE data, we further correct for the systematic offsets in $w_{\rm gmag}$ that lead to underestimation of the global galaxy fluxes based on the information provided by the WISE Explanatory Supplement\footnote{http://wise2.ipac.caltech.edu/docs/release/allsky/expsup/sec6\_3e.html}.

\begin{figure*}
\begin{center} 
\vspace*{0 cm} 
\hspace*{-1 cm} 
\includegraphics[angle=0, scale=0.6]{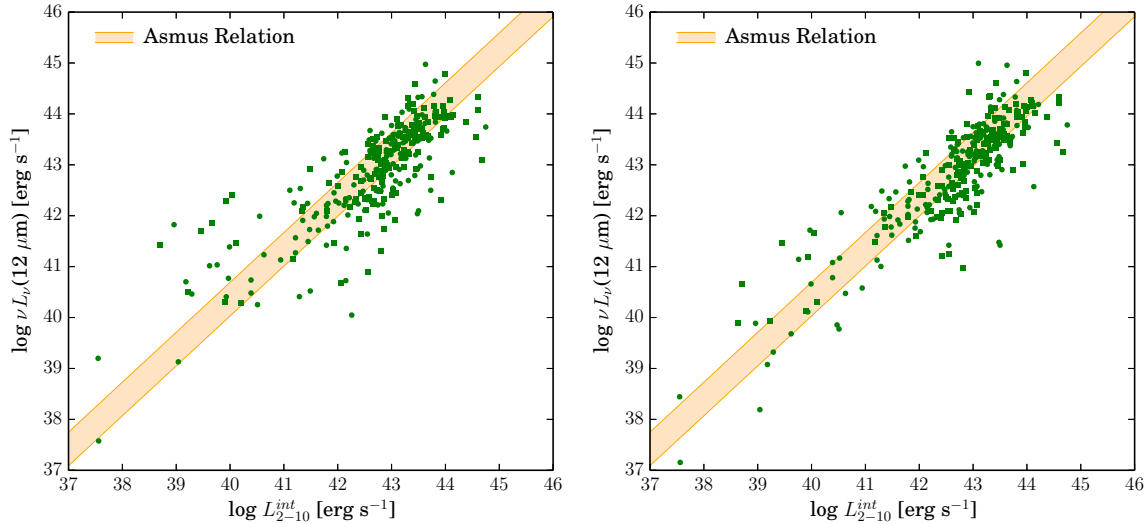}
\vspace*{0.0 cm} 
\caption{{\bf Left panel:}  The comparison between the intrinsic 2-10 keV luminosity $L_{2-10}^{int}$ and the 12 $\micron$ AGN luminosity $L_{\rm 12 \micron}^{AGN}$ $\equiv$ $\nu L_{\nu}$(12 $\micron$) [erg~s$^{-1}$], $L_{\rm 12 \micron}^{AGN}$ obtained from the SED-ftting based on global galaxy flux. The pale orange line indicates the correlation between $L_{2-10}^{int}$ and $L_{\rm 12 \micron}^{AGN}$ found by Asmus et al. (2015), with the line width showing the intrinsic scatter of the Asmus relation (0.33 dex). {\bf Right panel:} The same comparison with $L_{\rm 12 \micron}^{AGN}$ measured based on 6\arcsec aperture photometry in the UV-near-IR bands and profile-fit magnitudes in the WISE bands. The intrinsic scatters with respect to the Asmus relation 
are 0.58 dex and 0.82 dex for sources having $L_{2-10}^{int}$ $>$ 10$^{41}$ erg~s$^{-1}$ and $L_{2-10}^{int}$ $<$ 10$^{41}$ erg~s$^{-1}$, respectively.  }
\label{xray-mir}
\end{center} 
\end{figure*}

\subsubsection{Host Galaxy Contamination in the mid-IR AGN luminosity}
Before using $L^{AGN}_{12 \micron}$ as measured from SED-fitting to infer the absorbing column density using comparisons with $L^{obs}_{2-10}$, it is important to check whether the residual mid-IR emission from star-forming activity (hereafter referred to as host galaxy contamination) contributes substantially to $L_{\rm 12 \micron}^{AGN}$. Significant host galaxy contamination in the mid-IR AGN luminosity will make the ratio $L^{obs}_{2-10}$/$L_{\rm 12 \micron}^{AGN}$ artificially low and result in an overestimate of the absorbing column density. 

To determine the level of any host galaxy contamination, we compare $L^{AGN}_{\rm 12 \micron}$ as measured from the SED fitting to the prediction based on the {\it Asmus relation} (Asmus et al. 2015), which is the extension of the Gandhi relation to AGN having a low luminosity of $L_{2-10}^{int}$ $\sim$10$^{39}$ erg~s$^{-1}$. In order to make predictions with the Asmus relation, we require reliable measurements of $L_{2-10}^{int}$ (i.e., those that aren't affected by the systematic biases caused by heavy absorbing columns in Compton-thick AGN).
For this comparison we therefore primarily use galaxies from the {\it Swift}/BAT catalog.  We also include, where appropriate, Compton-thin sources from the {\it XMM-Newton} catalog, for which $L_{2-10}^{int}$ can be reliably inferred from the best-fit parameters of the X-ray spectral analysis in XMMFITCAT. 

To differentiate between Compton-thin and Compton-thick sources, we first collect the {\it XMM-Newton} galaxies that have X-ray-to-[OIII] luminosity ratios (i.e. the $T$-ratio; $T$ $\equiv$ $L_{2-10}^{obs}$/$L_{\rm [OIII]}$) available, followed by examining these sources with the $T$-ratio method from Bassani et al. (1999).  The $T$-ratio method provides a convenient way to separate Compton-thin AGN ($T$ $>$ 1) from Compton-thick sources ($T$ $<$ 1; Bassani et al. 1999; Cappi et al. 2006; Panessa et al. 2006). Considering the effect of X-ray variability and the uncertainty in the reddening correction to $L_{\rm [OIII]}$ (Brightman \& Nandra 2011), we set $L_{2-10}^{obs}$/$L_{\rm [OIII]}$ $>$ 5 to better select Compton-thin AGN in our comparison for {\it XMM-Newton} sources. Note that the [OIII] luminosities used in our examination were gathered via an extensive literature search (see references in Table 1); of all the reported $L_{\rm [OIII]}$ values, we only choose those measurements having a Balmer decrement available for reddening correction, for which we assume an intrinsic ratio of (H$\alpha$/H$\beta$)$_{0}$ of 3.0 and a H$\beta$/H$\alpha$ color index for extinction of 2.94 (Bassani et al. 1999).

Figure \ref{xray-mir} presents the relationship between $L_{2-10}^{int}$ and $L_{\rm 12 \micron}^{AGN}$ $\equiv$ $\nu L_{\nu}$(12 $\micron$) based on two different sets of flux measurements, along with the Asmus relation.   
The first set of measurements (left panel) employs global galaxy fluxes for the SED fitting. While these measurements are generally consistent with the Asmus relation for the bright X-ray AGNs (log $L_{2-10}^{int}\gtrsim 42.5$), they do not exhibit any apparent correlation below this threshold.  
This behavior matches the findings of Mullaney et al. (2011), who showed that $L_{\rm MIR}$ derived from SED fitting might include significant host galaxy contamination for lower luminosity AGNs.  Therefore, in order to infer reliable absorbing column densities for the lower luminosity sources, we need to find ways to obtain more accurate $L_{\rm 12 \micron}^{AGN}$ estimates, so as to substantially reduce the scatter in the correlation with $L_{2-10}^{int}$.

\subsubsection{The Approach to Mimimizing the Host Galaxy Contamination}
To minimize the host galaxy contamination in lower luminosity AGN while retaining the total AGN flux, we find -- for all WISE bands -- that it is helpful for the SED fitting to use the profile-fit magnitudes (w$_{\rm mpro}$) rather than using w$_{\rm gmag}$.  Because AGNs are point sources at {\it WISE} resolutions, the profile-fit magnitudes are more suitable as they capture all of the AGN flux while excluding mid-IR emission from star-forming activities outside of the PSF (for spatially resolved sources). 

To minimize the host galaxy contribution in the optical and near-IR bands, we tested SED fits using fluxes obtained from apertures smaller than 16.6 arcsec, which correspons to the mean half-light radius of
the X-ray sample galaxies. We choose the aperture size to be 6\arcsec, which matches the {\it WISE} resolutions at W1, W2 bands. For the SDSS data, we calculate the 6\arcsec aperture photometry by first determining the surface brightness model (i.e. the {\it de Vaucouleurs} or the {\it exponential} profile) used to compute the SDSS ``model'' magnitude based on the fitting likelihood for each model, followed by inferring the 6\arcsec aperture magnitude with the best-fit model and effective radius as listed in the SDSS catalog. For the near-IR data, we infer the 6\arcsec photometry from the 2MASS curve-of-growth determined using aperture sizes of 5\arcsec, 7\arcsec, 10\arcsec, 15\arcsec, and 20\arcsec. 

The right panel of Figure \ref{xray-mir} presents the $L^{AGN}_{12 \micron}$ values obtained from fitting these host-galaxy-minimized SED measurements.
We see that these $L_{\rm MIR}$ match the prediction of the Asmus relation across the six orders of magnitude spanned by our sample.  We also see that tthe scatter in the $L^{AGN}_{12 \micron}-L^{int}_{2-10}$ relation has been substantially reduced for the data with log$L_{2-10}^{int}$ less than 42.5.  This behavior suggests that the 12 $\micron$ AGN luminosity derived from the aperture photometry SED fitting is reliable and can be used to build the $L^{AGN}_{12 \micron} - L^{obs}_{2-10}$ diagram for investigating the maser detection problem.  Note that despite the relatively large scatter (0.82 dex) still remaining for the log$L_{2-10}^{int} < 41$ data, the impact on our conclusions will remain minimal as nearly all maser galaxies have $L_{2-10}^{int}$ greater than $10^{41}$ erg~s$^{-1}$.



\section{Results : Consequences for the megamaser detection rates}

Before looking for ways to enhance the maser detection rates using the mid-IR and X-ray photometry simultaneously, it is useful to first study the dependence of maser detection rates on the mid-IR AGN luminosity ($L^{AGN}_{\rm 12 \micron}$) and absorbing column density ($N_{\rm H}$), both separately and in concert.  Evaluating the relative importance of these two factors may help to identify sample selection criteria that simultaneously maximize both the detection and completion rates for maser galaxies. 

\subsection{The Effect of mid-IR AGN Luminosity}

\begin{figure*}
\begin{center} 
\vspace*{-1.5 cm} 
\hspace*{-1 cm} 
\includegraphics[angle=0, scale=0.6]{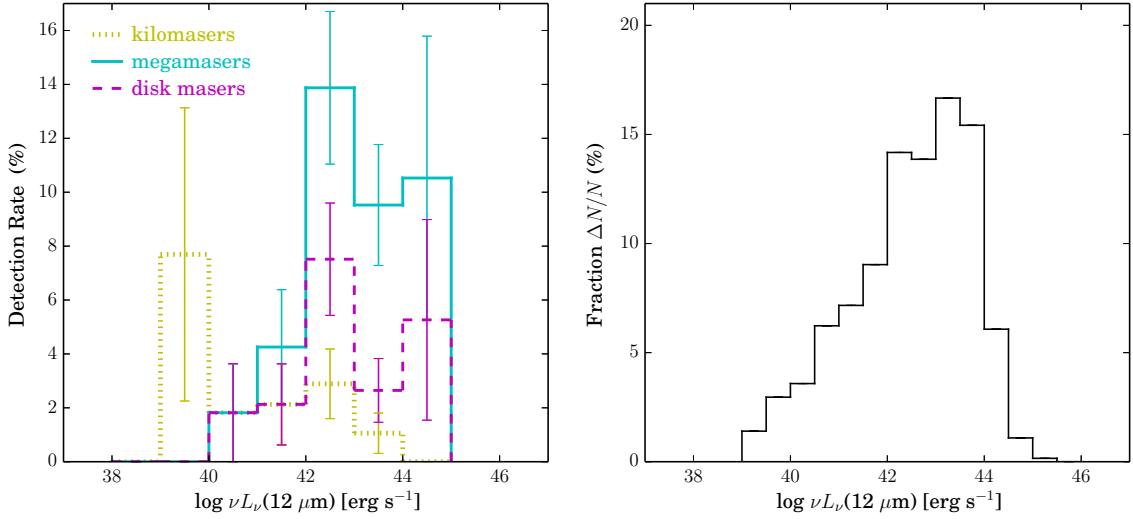}
\vspace*{0.0 cm} 
\caption{{\bf Left panel:} The detection rates of kilomasers (dotted yellow), megamasers (continue cyan), and disk masers (dashed magenta) as functions of $L^{AGN}_{\rm 12 \micron}$.  {\bf Right panel:} The fraction of galaxies in the X-ray sample as a function of 12 $\micron$ AGN luminosity $L^{AGN}_{\rm 12 \micron}$. $\Delta N$ and $N$ refer to the number of sources in each luminosity bin and the total number of galaxies in the X-ray sample, respectively.   }
\label{mir-stat}
\end{center} 
\end{figure*}

In the left panel of Figure \ref{mir-stat}, we plot the detection rates of kilomasers (dotted yellow), megamasers (solid cyan), and disk masers (dashed magenta) for the X-ray sample. From this plot, we see that the detection rates of megamasers and disk masers first increase steadily as the 12 $\micron$ AGN luminosity rises from 10$^{38}$ erg~s$^{-1}$ to 10$^{42}$ erg~s$^{-1}$ and then get boosted substantially by a factor of $\sim$2.7 when $L^{AGN}_{\rm 12 \micron}$ is greater than 10$^{42}$ erg~s$^{-1}$ (i.e. $L^{int}_{2-10}$ $\ge$ 5.0$\times$10$^{41}$ erg~s$^{-1}$, according to the Asmus relation). This behavior suggests that, even without including the effect of $N_{\rm H}$, one could immediately increase the megamaser detection rates by a factor of $\sim$3 relative to the GBT sample\footnote{The detection rates of all masers, megamasers, and disk masers in the GBT sample are 3.3$\pm$0.1\%, 2.7$\pm$1.1\%, and 0.9$\pm$0.8\%, respectively (see Table 3 in Kuo et al. 2018).} by making a mid-IR AGN luminosity cut of $L^{AGN}_{\rm 12 \micron}$ $\ge$ 10$^{42}$ erg~s$^{-1}$ when selecting Type II AGNs for maser surveys.

However, as shown in the right panel of Figure \ref{mir-stat}, the majority (67\%) of galaxies in the X-ray sample already have $L^{AGN}_{\rm 12 \micron}$ $\ge$ 10$^{42}$ erg~s$^{-1}$. The galaxy distribution is biased towards higher mid-IR AGN luminosity by the inclusion of the {\it Swift}/BAT sample, which primarily consists of sources with $L^{AGN}_{\rm 12 \micron}$ $>$ 10$^{42}$ erg~s$^{-1}$.
Since the X-ray sample is dominated by these higher luminosity AGN, making the mid-IR AGN luminosity cut of $L^{AGN}_{\rm 12 \micron}$ $\ge$ 10$^{42}$ erg~s$^{-1}$ in the X-ray sample only leads to an increase of megamaser detection rates by a factor of $\sim$1.4 with respect to the X-ray sample itself. 

\subsection{The Effect of the Obscuring Column Density }
To explore how the maser detection rates depend on obscuring column density, we first infer $N_{\rm H}$ for the X-ray sample from the ratio of the observed X-ray to mid-IR AGN luminosity based on the $L^{AGN}_{12 \micron}$$-$$L^{obs}_{2-10}$ diagram (Satyapal et al. 2017). This diagram allows one to estimate the $N_{\rm H}$ value of an AGN because the more obscured an AGN is, the stronger the X-ray absorption will be, suppressing the observed X-ray luminosity $L^{obs}_{2-10}$. Since the reprocessed mid-IR continuum emission in AGN is relatively unaffected by the X-ray absorbing material and thus gives a reliable estimate of the intrinsic AGN luminosity (even for Compton thick sources; Goulding et al. 2011), the ratio of the observed X-ray to mid-IR AGN luminosity 
allows one to evaluate the magnitude of suppression in $L^{obs}_{2-10}$ and thus provides an estimate of $N_{\rm H}$, especially at $N_{\rm H}$ $>$ 10$^{23}$ cm$^{-2}$ (Asmus et al. 2015).

\begin{figure*}
\begin{center} 
\vspace*{0 cm} 
\hspace*{0.0 cm} 
\includegraphics[angle=0, scale=0.75]{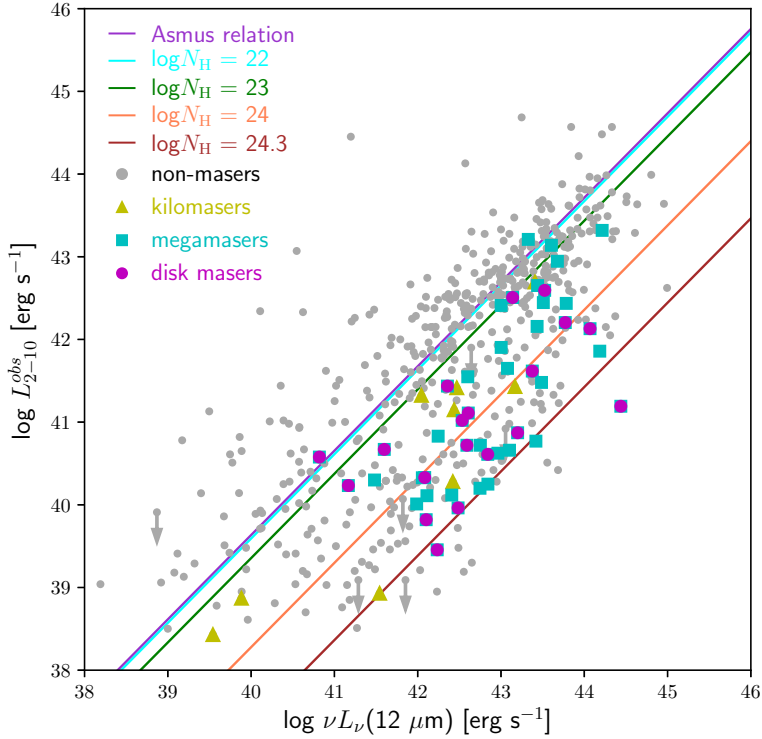}
\vspace*{0.0 cm} 
\caption{ The $L^{AGN}_{12 \micron}-L^{obs}_{2-10}$ diagram. The horizontal axis shows the 12 $\micron$ AGN luminosity $L^{AGN}_{12 \micron}$ obtained from the SED fitting. The vertical axis indicates the observed 2-10 keV X-ray luminosity $L^{obs}_{2-10}$. The kilomasers are shown as filled yellow triangles, megamasers as cyan squares, disks as magenta filled circles, and non-masers as small grey circles. The purple line represents the Asmus relation. The cyan, green, orange and brown lines indicate the observed 2-10 keV luminosity expected for sources with log$N_{\rm H}$ $=$ 22, 23, 24, and 24.3 cm$^{-2}$, respectively.}
\label{asmus_gbt}
\end{center} 
\end{figure*}

In Figure \ref{asmus_gbt}, we plot $L^{AGN}_{\rm 12 \micron}$ against $L^{obs}_{2-10}$ for the sources in the X-ray sample. In this diagram, the yellow triangles, cyan squares, and magenta filled circles represent the kilomasers, megamasers, and disk masers, respectively; the sample of X-ray galaxies that have been searched for 22 GHz emission with the GBT but have no detection (i.e., the non-masers) is shown as grey bullets. 

To estimate the $N_{\rm H}$ value of each AGN, we compare $L_{2-10}^{obs}$ with $L_{2-10}^{int}$ evaluated from the Asmus relation (the purple line) at each given $L_{\rm 12 \micron}^{AGN}$. The cyan, green, orange, and brown lines indicate the $L^{obs}_{2-10}$ expected for sources with log$N_{\rm H}$ $=$ 22, 23, 24, and 24.3 cm$^{-2}$, respectively, based\footnote{log$L_{2-10}^{obs}$ $=$ log$L_{2-10}^{int}$ $-$ $\epsilon$, where $\epsilon$ $=$ 0.0414, 0.2810, 1.3556, and 2.2900 for log$N_{\rm H}$ $=$ 22, 23, 24, and 24.3 cm$^{-2}$, respectively. } on Equation (1) and the assumption that the photon index $\Gamma=1.8$ (Nandra \& Pounds 1994; Dadina 2008). We note that while we do not include a reflected X-ray component from the AGN (which becomes important only when log$N_{\rm H}$ $\gtrsim$ 24.3), our prediction is nevertheless consistent with that made by the MYTORUS model\footnote{log$L_{2-10}^{obs}$ $=$ log$L_{2-10}^{int}$ $-$ $\eta$, where $\eta$ $=$ 0.2929, 1.3673, and 2.20703 for log$N_{\rm H}$ $=$ 23, 24, and 24.3 cm$^{-2}$, respectively. } (Murphy \& Yaqoob 2009; Satyapal et al. 2017), which considers
both intrinsic absorption and reprocessed X-ray emission from a toroidal absorber for the full range of column density distributions. Therefore, replacing our prediction with that from the MYTORUS model does not change our conclusion.

The comparison shown in Figure \ref{asmus_gbt} indicates that the majority (93\%) of maser sources have $N_{\rm H}$ $\gtrsim$ 10$^{23}$ cm$^{-2}$, consistent with the findings from Greenhill et al. (2008) and Castangia et al. (2013) who derived the $N_{\rm H}$ of maser sources based on X-ray spectral fitting.  This behavior suggests that the maser detection rates are strong functions of absorbing column density.

In the first four rows of Table 4, we list the detection rates for all masers, megamasers, and disk masers for sources having $N_{\rm H}$ $<$ 10$^{23}$ cm$^{-2}$, 10$^{23}$ $<$ $N_{\rm H}$ $<$ 10$^{24}$ cm$^{-2}$, $N_{\rm H}$ $\ge$ 10$^{23}$ cm$^{-2}$, and $N_{\rm H}$ $\ge$ 10$^{24}$ cm$^{-2}$, respectively. The table shows that the maser detection rates get boosted by more than an order of magnitude when the column density of an AGN increases from $N_{\rm H}$ $<$ 10$^{23}$ cm$^{-2}$ to $N_{\rm H}$ $\ge$ 10$^{23}$ cm$^{-2}$. Moreover, with respect to the maser detection rates of the GBT sample, this table also shows that one can increase the detection rate of megamasers (disk masers) by a factor of $\sim$6 (7) and $\sim$7 (10) by choosing galaxies with $N_{\rm H}$ $\ge$ 10$^{23}$ cm$^{-2}$ and $N_{\rm H}$ $\ge$ 10$^{24}$ cm$^{-2}$, respectively.

Interestingly, unlike sample selections based solely on the mid-IR colors (Kuo et al. 2018), these dramatic increases in maser detection rates do not come at the cost of the completeness rates (C$_{\rm maser}$) when X-ray detection and information is included.  Here, C$_{\rm maser}$ is defined as $N^{\rm cut}_{\rm maser}$/$N^{\rm tot}_{\rm maser}$, where $N^{\rm tot}_{\rm maser}$ indicates the total number of maser detections in the X-ray sample and $N^{\rm cut}_{\rm maser}$ refers to the number of detected masers which satisfy certain cuts in $L^{AGN}_{12 \micron}$ or $N_{\rm H}$. When evaluating the completeness rates for megamasers (C$_{\rm Mmaser}$) or disk masers (C$_{\rm disk}$), we replace the  $N^{\rm cut}_{\rm maser}$ in the definition for C$_{\rm maser}$ by $N^{\rm cut}_{\rm Mmaser}$ and $N^{\rm cut}_{\rm disk }$, where $N^{\rm cut}_{\rm Mmaser}$ and $N^{\rm cut}_{\rm disk }$ indicates the number of detected megamasers and disk masers which meet the chosen selection criteria, respectively.

From Table 1, it can be seen that the X-ray sample completeness rate for sources with $N_{\rm H}$ $\ge$ 10$^{23}$ cm$^{-2}$ remains close to 100\% for megamasers (C$_{\rm Mmaser}$ $=$ 93.3\%) and disk masers (C$_{\rm disk}$ $=$ 94.7\%), even when the detection rates are as high as 18.8$\pm$2.6\% and 6.4$\pm$1.5\%, respectively. Even if one wants to achieve the highest the maser detection rates (i.e. 25.0$\pm$4.5\%) by selecting Compton-thick candidates (i.e. $N_{\rm H}$ $\ge$ 10$^{24}$ cm$^{-2}$), the completeness rates ($\sim$50\%) remain substantially higher than the rates ($\sim$30\%) achieved by the mid-IR selection criterion of $W1-W2>0.5$ \& $W1-W4>7$ that led to the highest detection rate (18\%) in Kuo et al. (2018). This result indicates that selecting sources based on the obscuring column density allows one to achieve both high detection rates and high completeness rates simultaneously.

  \begin{deluxetable*}{lrcccccccc} 
\tabletypesize{\scriptsize} 
\tablewidth{0 pt} 
\tablecaption{Effective Search Criteria for Megamasers \& Disks \label{tbl-summary}}
\tablehead{ 
\colhead{log~$N_{\rm H}$ } & \colhead{$L^{AGN}_{\rm 12 \micron}$ } & \colhead{\% All MCP} & \colhead{$R_{\rm maser}$} & \colhead{C$_{\rm maser}$} & \colhead{$R_{\rm Mmaser}$} & \colhead{C$_{\rm Mmaser}$} & \colhead{$R_{\rm disk}$} & \colhead{C$_{\rm disk}$}  & \colhead{$N_{\rm disk}$}  \\
\colhead{(cm$^{-2}$)}  & \colhead{(erg~s$^{-1}$)}      & \colhead{Galaxies}   & \colhead{\%} & \colhead{\%} & \colhead{\%} & \colhead{\%} & \colhead{\%} & \colhead{\%}  & \colhead{ }}     
\startdata 
&  &  &  &  &  &  &  & &  \\
log~$N_{\rm H}$ $<$ 23 &  $>$ 10$^{38}$ & 48.7  & 1.1$\pm$0.7 & 5.4 & 1.1$\pm$0.7 &  6.7 &  0.4$\pm$0.4 & 5.3  &  1$\pm$1 \\
&  &  &  &  &  &  & & & \\\hline
&  &  &  &  &  &  &  & & \\
23 $\le$ log~$N_{\rm H}$ $<$ 24 &   $>$ 10$^{38}$  & 31.2  & 15.4$\pm$3.0 & 46.4 & 11.8$\pm$2.6 &  44.4 &  4.7$\pm$1.7 & 42.1  & 6$\pm$2 \\
&  &  &  &  &  &  & & & \\\hline
&  &  &  &  &  &  &  & &\\ 
{\bf log~$N_{\rm H}$ $\ge$ 23} &  {\bf $>$ 10$^{38}$} & {\bf51.2} & {\bf19.1$\pm$2.6} & {\bf94.6} & {\bf15.1$\pm$2.3} &  {\bf93.3} &  {\bf6.5$\pm$1.5} & {\bf94.7}  & {\bf 15$\pm$3}\\
&  &  &  &  &  &  & & & \\\hline
&  &  &  &  &  &  &  & &  \\ 
{\bf log~$N_{\rm H}$ $\ge$ 24} & {\bf $>$ 10$^{38}$ } & {\bf20.0} & {\bf25.0$\pm$4.8} & {\bf47.4} & {\bf20.3$\pm$4.3} &  {\bf48.9} &  {\bf9.3$\pm$2.9} & {\bf52.6} &  {\bf 9$\pm$3} \\
&  &  &  &  &  &  & & & \\\hline
&  &  &  &  &  &  &  &  &\\
log~$N_{\rm H}$ $<$ 23 &  $>$ 10$^{42}$ & 32.5 & 1.1$\pm$0.8 & 3.6 & 1.1$\pm$0.8 &  4.4 &  0.0$\pm$0.0 & 0.0 & 0\\
&  &  &  &  &  &  & & & \\\hline
&  &  &  &  &  &  &  & &\\
23 $\le$ log~$N_{\rm H}$ $<$ 24 &   $>$ 10$^{42}$ & 23.1  & 16.8$\pm$3.7 & 37.5 & 13.6$\pm$3.3 &  37.8 &  4.8$\pm$2.0 & 31.6  & 3$\pm$1 \\
&  &  &  &  &  &  & & & \\\hline
&  &  &  &  &  &  &  &  &\\ 
{\bf log~$N_{\rm H}$ $\ge$ 23 } & {\bf $>$ 10$^{42}$ } & {\bf 36.8} & {\bf 22.1$\pm$3.3} & {\bf 78.6} & {\bf 19.1$\pm$3.1} &  {\bf 84.4} &  {\bf 8.0$\pm$2.0} & {\bf 84.2}  & {\bf 10$\pm$3}\\
&  &  &  &  &  &  & & & \\\hline
&  &  &  &  &  &  &  &  &\\ 
{\bf log~$N_{\rm H}$ $\ge$ 24} & {\bf  $>$ 10$^{42}$} & {\bf 13.7} & {\bf 31.1$\pm$6.5} & {\bf 41.1} & {\bf 27.0$\pm$6.2} &  {\bf 46.7} &  {\bf 13.5$\pm$4.3} & {\bf 52.6 }  & {\bf 8$\pm$3}\\
&  &  &  &  &  &  & & & \\\hline
\enddata 
\tablecomments{ $R_{\rm maser}$, $R_{\rm Mmaser}$, and $R_{\rm disk}$ are the detection rates of all masers, megamasers, and disks, respectively.   C is the completeness rate, listed separately for all masers, megamasers, and disks.  The criteria that give the highest detection and completion rates for megamasers and disks are highlighted in bold. $N_{\rm disk}$ shows the predicted number of new disk masers that could be detected from sources in the {\it Swift}/BAT AGN catalog and the {\it XMM-Newton} spectral-fit database which are not yet searched for masers with the GBT. }
\end{deluxetable*}

\subsection{The Joint Effect of $L^{AGN}_{\rm 12 \micron}$ and $N_{\rm H}$}

Given that both the mid-IR AGN luminosity and the absorbing column density can play substantial roles for maser detection when examined individually, it is natural to explore whether combining the two factors could provide an even more effective optimization.  In the last four rows of Table 4, we show the maser detection rates and completeness rates as a function of log$N_{\rm H}$ for all sources having $L^{AGN}_{\rm 12 \micron}$ $\ge$ 10$^{42}$ erg~s$^{-1}$. We see that the detection rates of megamasers and disk masers in highly obscured and Compton-thick AGN increase by another factor of 1.3$-$1.5 by enforcing that the 12 $\micron$ AGN luminosity be greater than 10$^{42}$ erg~s$^{-1}$. While the mid-IR AGN luminosity appears to play a secondary role in boosting the megamaser detection rates, the highest detection rates achieved by making a cut in  $L^{AGN}_{\rm 12 \micron}$ become 27.0$\pm$6.2\% and 13.5$\pm$4.3\% for megamasers and disk masers, respectively, corresponding to a factor of 10 and 15 boost in the detection rates with respect to the GBT sample. Meanwhile, the completeness rates remain at the level of $\sim$80\%($\sim$50\%) for sources with $N_{\rm H}$ $\ge$ 10$^{23}$ cm$^{-2}$ ($N_{\rm H}$ $\ge$ 10$^{24}$ cm$^{-2}$), indicating that selecting AGN based on $L_{AGN}$ and $N_{\rm H}$ can lead to high maser detection rates and completeness rates simultaneously.


\begin{figure*}
\begin{center} 
\vspace*{0 cm} 
\hspace*{-0.5 cm} 
\includegraphics[angle=0, scale=0.55]{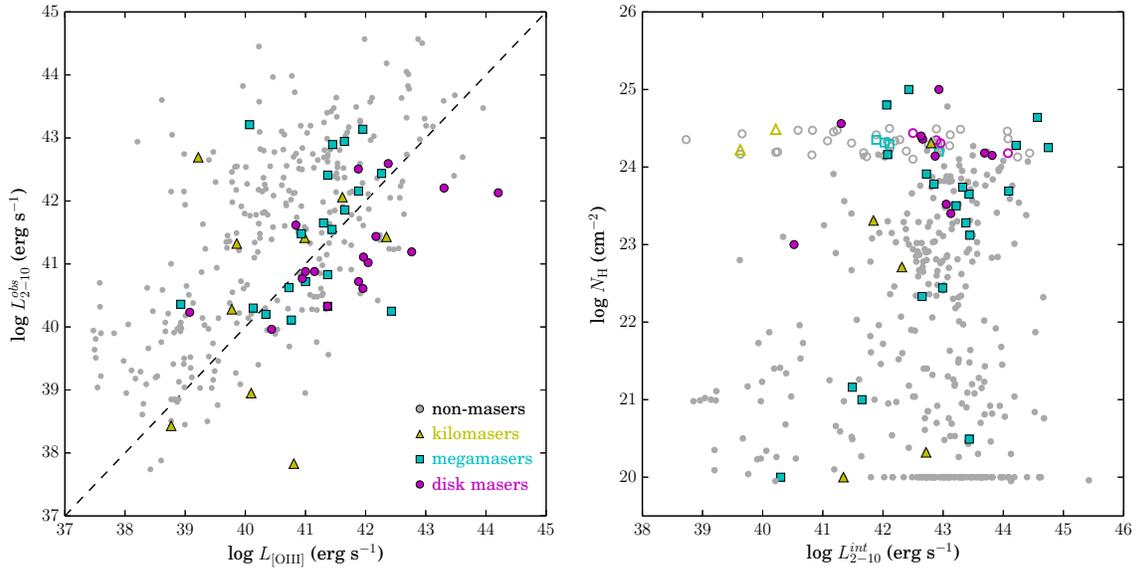}
\vspace*{0.0 cm} 
\caption{{\bf Left panel:} The $L_{\rm [OIII]}$$-$$L_{\rm X}$ diagram. The horizontal axis shows the reddening-corrected [OIII] $\lambda$5007 luminosity $L_{\rm [OIII]}$ whereas the vertical axis represent the observed 2-10 keV X-ray luminosity $L_{2-10}^{obs}$. The dashed line indicates the region where $L_{\rm [OIII]}$ $=$ $L_{2-10}^{obs}$. The sources below the dashed line have $T$ $\equiv$ $L_{2-10}^{obs}$/$L_{\rm [OIII]}$ $<$ 1, suggesting that they are candidates of Compton-thick AGN. {\bf Right panel:}  The distribution of the X-ray sample galaxies in the ($L_{2-10}^{int}$, $N_{\rm H}$) parameter space. The open circles indicates the absorbing column densities of the non-{\it Swift}/BAT sources which have $T$-ratios smaller than one. For these sources, we follow Cappi et al. (2006) to set log$N_{\rm H}$ $=$ 24.3. To avoid significant clustering and overlapping of points, we randomize log$N_{\rm H}$ such that the column densities are uniformly distributed between 24.1 and 24.5.  At the bottom of the figure, there exists an apparent floor at log$N_{\rm H}$ $=$ 20 because this is lower limit of log$N_{\rm H}$ in the X-ray spectral fitting and the $N_{\rm H}$ value around which galactic absorption along the line of sight begins to dominate the absorbing column. }
\label{lo3-lx-nh}
\end{center} 
\end{figure*}

\subsection{The $L^{\rm obs}_{2-10}$--$L_{\rm [OIII]}$ Diagram}

To examine the robustness of our conclusion made in the previous section, which is drawn from an indirect estimate of absorbing column density based on the $L^{obs}_{2-10}$$-$$L^{AGN}_{12 \micron}$ diagram, it would be useful to see if we can arrive at the same conclusion by using independent estimates or direct measurements of $N_{\rm H}$, which are available for subsets of the X-ray sample. 

Among the 642 sources in the X-ray sample, 342 (53\%) of them have reddening-corrected [OIII] $\lambda$5007 luminosities $L_{\rm [OIII]}$ available in the literature (see Table 2). Similar to the reprocessed mid-IR continuum emission, $L_{\rm [OIII]}$ is often used as an indicator of the isotropic AGN luminosity because it is not affected by the optically thick X-ray absorbing material (Heckman et al. 2005; Panessa et al. 2006; Goulding et al. 2011; Koulouridis et al. 2016). Through the examination of a local sample of Seyfert 2 galaxies, Bassani et al. (1999) demonstrated that the X-ray$-$[OIII] flux ratio (i.e. the T-ratio; $T$ $\equiv$ $L_{2-10}^{obs}$/$L_{\rm [OIII]}$) of $T$ $<$ 1 provides a reliable diagnostic for Compton-thick AGN. Therefore, for the subset of the X-ray sample with $L_{\rm [OIII]}$ available, we are able to explore the maser detection rates for Compton-thick sources by identifying these sources based on the $T$-ratio method.

In the left panel of Figure \ref{lo3-lx-nh}, we plot $L_{\rm [OIII]}$ against the observed 2-10 keV AGN luminosity.  The dashed line indicates the region where the T-ratio equals 1. All of the sources lying below the dashed lines (i.e. $T$ $<$ 1) are candidates for Compton-thick AGN. Based on the distribution seen in the plot, we can infer that the detection rates of all masers, megamasers, and disk masers in the Compton-thick candidates are 25.5$\pm$5.2\%, 21.3$\pm$4.8\%, 13.8$\pm$3.8\%, respectively. These values are consistent with the maser detection rates inferred from the $L^{AGN}_{12 \micron}$$-$$L^{obs}_{2-10}$ diagram for sources with log$N_{\rm H}$ $\ge$ 24 (see the 4th row of Table 4), supporting the conclusions from Section 3.2 for Compton-thick sources.

Note that, although $L_{\rm [OIII]}$ provides an independent measurement of the intrinsic AGN luminosity, we do not use the the ratio of the observed X-ray to [OIII] luminosity to infer the absorbing column density as we did in Section 3.2 for Compton-thin sources (i.e. log$N_{\rm H}$ $<$ 24). This is because the intrinsic 2-10 keV luminosity and $L_{\rm [OIII]}$ do not show a tight correlation (Heckman et al. 2005; Georgantopoulos \& Akylas 2010; Berney et al. 2015), particularly for Seyfert 2 galaxies. Nevertheless, one can see from the left panel of Figure \ref{lo3-lx-nh} that some of maser galaxies do have $T$ $>$ 1. This suggests that maser galaxies reside in Compton-thin AGN as well, consistent with what we see in Figure \ref{asmus_gbt}. To robustly confirm the maser detection rates in Compton-thin sources, it would be best to use a direct measurement of $N_{\rm H}$ from X-ray spectral fitting.

\subsection{The L$^{int}_{2-10}-N_{\rm H}$ Diagram}

The spectral analyses from the X-ray surveys provide the absorbing column densities $N_{\rm H}$ for all sources in the X-ray sample (see Table 2). However, out of the 314 X-ray sources from the {\it XMM-Newton} catalog, there are 132 sources for which the column densities were fixed in the spectral-fitting process. This happens when the quality of the spectrum is not high enough (e.g. low signal-to-noise ratio) to fit $N_{\rm H}$ robustly. These $N_{\rm H}$ values tend to be subject to greater uncertainties and we do not use them in our present comparison.

Furthermore, for sources drawn from X-ray surveys probing $\le$10 keV bands (e.g.  {\it XMM-Newton} and {\it Chandra}), the $N_{\rm H}$ values can be significantly underestimated if the AGN are Compton-thick (e.g. Cappi et al. 2006; Panessa et al. 2006; Singh, Shastri, \& Risaliti 2011; Castangia et al. 2013). Therefore, to reliably measure the maser detection rates as a function of $N_{\rm H}$, we primarily use galaxies from the {\it Swift}/BAT AGN catalog, which provides reliable $N_{\rm H}$ even for Compton-thick sources. In addition, we also use {\it XMM-Newton} sources that have $L_{\rm [OIII]}$ available so that we can infer a more reliable $N_{\rm H}$ for Compton-thick AGN based on the $T$-ratio method. In total, this sample of more reliable $N_{\rm H}$ measurements includes 417 sources, with 307 from the {\it Swift}/BAT AGN catalog and 110 from XMMFITCAT.
 
\begin{figure*}
\begin{center} 
\vspace*{0 cm} 
\hspace*{-0.5 cm} 
\includegraphics[angle=0, scale=0.55]{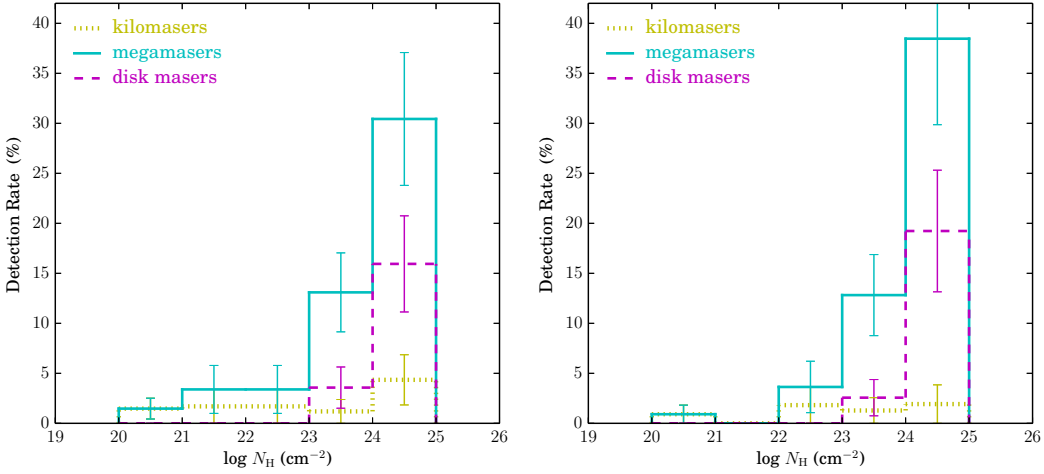}
\vspace*{0.0 cm} 
\caption{{\bf Left panel:} The detection rates of kilomasers, megamasers, and disk masers as a functions of log$N_{\rm H}$ for the entire X-sample galaxies, with $N_{\rm H}$ directly measured from X-ray spectral fitting.  {\bf Right panel:}  The same detection distributions plotted only for sources with $L_{\rm 12 \micron}$ $\ge$ 10$^{42}$ erg~s$^{-1}$.  }
\label{nh-stat}
\end{center} 
\end{figure*}

In the right panel of Figure \ref{lo3-lx-nh}, we plot $N_{\rm H}$ as a function of $L_{2-10}^{int}$. In this diagram, for those sources drawn from the {\it XMM-Newton} catalog with $T$ $<$ 1, we follow Cappi et al. (2006) in setting log$N_{\rm H}$ $=$ 24.3 to indicate that these sources are Compton-thick\footnote{If log$N_{\rm H}$ equals 24.3, it indicates that 99.5\% of the AGN X-ray emission is blocked by the obscuring material, assuming the photon index of the X-ray spectrum to be $\Gamma$ $=$ 1.8.}. We do not use the $N_{\rm H}$ provided by XMMFITCAT for these sources because their $N_{\rm H}$ are likely to be underestimated. To avoid significant clustering and overlapping of points in the figure, we randomize the log$N_{\rm H}$ values uniformly between 24.1 and 24.5. The inferred maser detection rates as a function of log$N_{\rm H}$ are shown in the left panel of Figure \ref{nh-stat}. From this plot, we can see that the overall distribution is consistent with what we show in Table 4. To further compare the detection rate distributions for more luminous sources, we show in the right panel of Figure \ref{nh-stat} the maser detection rates as a function of log$N_{\rm H}$ for sources with log$L_{2-10}^{int}$ $\ge$ 41.7 (i.e. $L^{AGN}_{\rm 12 \micron}$ $\ge$ 10$^{42}$ erg~s$^{-1}$). 

While the distribution remains unchanged for log$N_{\rm H}$ $<$ 23, the detection rates increase from 30.4$\pm$6.6\% to 38.5$\pm$8.6\% for megamasers and from 15.9$\pm$4.8\% to 19.2$\pm$6.2\% for disk masers. Although there are 11.5\% and 5.7\% increases relative to the detection rates for Compton-thick megamasers and disk masers shown in Table 4, respectively, the rates are consistent at the $\sim$1 $\sigma$ level.



\subsection{Photon Index and Eddington Ratio}
\begin{figure*}
\begin{center} 
\vspace*{0 cm} 
\hspace*{-3 cm} 
\includegraphics[angle=0, scale=0.43]{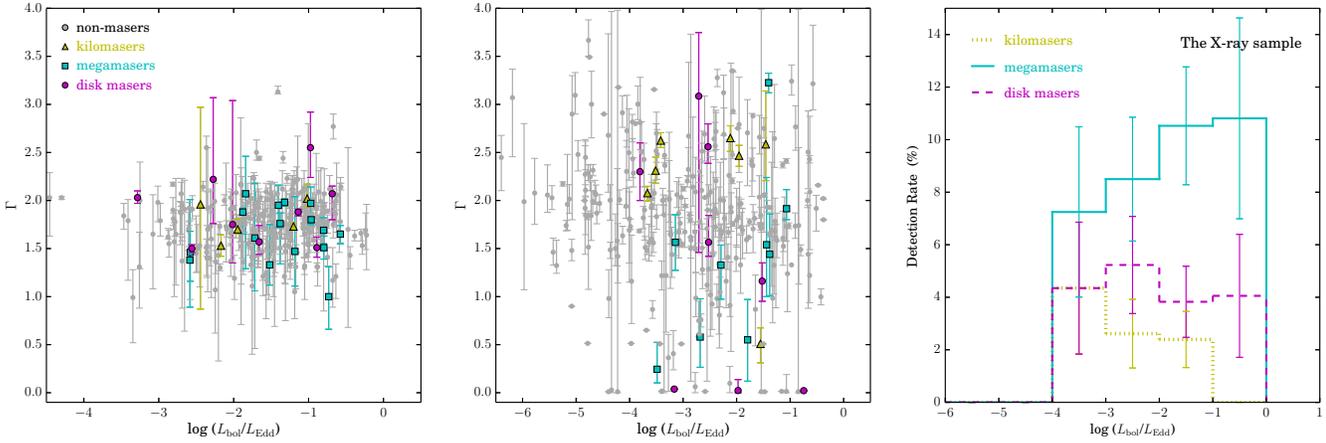}
\vspace*{0.0 cm} 
\caption{{\bf Left \& Middle panels:}  The distributions of the {\it Swift}/BAT (left) and  non-{\it Swift}/BAT (middle) sources of the X-ray sample in the log~($L_{\rm bol}$/$L_{\rm Edd}$) and $\Gamma$ parameter space. Here, $\Gamma$ refers to the photon index of an X-ray spectrum. {\bf Right panel:} The maser detection rates as a function of Eddington ratio.  The error bars illustrate the associated Poisson uncertainties. }
\label{edd-stat}
\end{center} 
\end{figure*}

We have shown that the maser detection rates depend substantially on two fundamental properties: the MIR intrinsic luminosity and the obscuring column density. Given the tight relationship between the maser phenomena and AGN activities, one may wonder whether the maser detection rates also correlate with another fundamental AGN property: the Eddington ratio $\lambda_{\rm Edd}$ $\equiv$ $L_{\rm bol}$/$L_{\rm Edd}$ (where $L_{\rm bol}$ and $L_{\rm Edd}$ refer to the bolometric and Eddington luminosity, respectively). Since $\lambda_{\rm Edd}$ is known to be correlated with the photon index $\Gamma$ of the AGN X-ray spectrum (e.g. Shemmer et al. 2006; Constantin et al. 2009; Brightman et al. 2013), it would be interesting to see whether masers are correlated with $\Gamma$ as well. We explore these possibilities in Figure \ref{edd-stat}.

The left and middle panels of Figure \ref{edd-stat} show the $\Gamma-\lambda_{\rm Edd}$ diagrams for sources from the {\it Swift}/BAT catalog and the non-{\it Swift}/BAT catalogs, respectively. We make the plots for the two different subsets of the X-ray sample separately because the photon indices provided by the {\it Swift}/BAT survey measure the slopes of the {\it intrinsic} power-law spectra, whereas $\Gamma$ values in the non-{\it Swift}/BAT catalog only indicate the {\it apparent} slopes of the X-ray spectra (which could contain significant reflection components in Compton-thick sources).

When evaluating $\lambda_{\rm Edd}$, we obtain $L_{\rm bol}$ from the 12 $\micron$ AGN luminosity by adopting the luminosity-dependent bolometric correction from Gandhi et al. (2009).  The Eddington luminosity is calculated from the black hole mass as :
\begin{equation}
L_{\rm Edd} = 1.25 \times 10^{38} \Big( { M_{\rm BH} \over M_{\odot} } \Big) ~{\rm erg~s^{-1} }~.
\end{equation}
Following Pons et al. (2016), we infer the black hole mass $M_{\rm BH}$ from the rest-frame total $K$-band luminosity $L_{K,\rm tot}$ of the host galaxy using the $M_{\rm BH}$-$L_{K,\rm tot}$ relation from L$\ddot{\rm a}$sker et al. (2014). Given the intrinsic scatter seen in the $M_{\rm BH}$-$L_{K,\rm tot}$ relation, the expected scatter in log~$L_{\rm Edd}$ is $\sim$0.5 dex. 

From the plot showing the {\it Swift}/BAT sample, we see that there is no clear correlation between the photon indices and the maser sources, suggesting that the slopes of the intrinsic AGN X-ray spectra for the maser galaxies are statistically similar.  In contrast, for sources in the non-{\it Swift}/BAT surveys, represented mainly by the {\it XMM-Newton} sources\footnote{Here the best-fit is obtained by the double power-law model, and we adopt the smaller of the two $\Gamma$ values provided by XMMFITCAT. The power-law spectrum associated with this photon index may reflect the scattered component of the hard X-ray emission (Corral et al. 2015).}, the middle panel of Figure \ref{edd-stat} shows that megamasers and disk masers preferentially have lower $\Gamma$ than the average value of the {\it Swift}/BAT sample (i.e. $\Gamma_{\rm avg} = 1.76$). This is consistent with the fact that $\sim$50\% of megamaser galaxies reside in AGN with $N_{\rm H}$ $\gtrsim$ 10$^{24}$ cm$^{-2}$, in which strong reflection components flatten the X-ray spectra at higher energies and give rise to smaller photon indices (e.g. Matt et al. 2000; Cappi et al. 2006; Koulouridis et al. 2016).  

By inspecting the maser distribution in the domain of Eddington ratio in the right panel of Figure \ref{edd-stat}, we see that the maser detection rates for sources having $\lambda_{\rm Edd}$ below 10$^{-4}$ are zero, suggesting that $\lambda_{\rm Edd}$ $\sim$ 10$^{-4}$  is an accretion efficiency threshold for maser excitation. We note, however, that this interpretation is based on a small fraction (11\%) of the X-ray sample sources with $\lambda_{\rm Edd}$ $<$ 10$^{-4}$, which may not be statistically significant enough to draw firm conclusion.  In the X-ray sample we employ in thi sanalysis, the sources with $\lambda_{\rm Edd}$ $<$ 10$^{-4}$ mainly come from the {\it XMM-Newton} catalog, while such low $\lambda_{\rm Edd}$ AGN are nearly absent in the {\it Swift}/BAT sample because the spectral analysis in the {\it Swift}/BAT catalog requires sufficient photon counts $N$ up to 150 keV, where $N$ can become significantly lower than the photon counts at $\le$10 keV bands; therefore, the {\it Swift}/BAT sample tends to include more luminous and higher $\lambda_{\rm Edd}$ AGN than the {\it XMM-Newton} sample in order to achieve sufficient sensitivity across the wide energy bands.  Thus, the robustness of this interpretation will need to be tested by including more low $\lambda_{\rm Edd}$ sources in the analysis in the future.

For sources with $\lambda_{\rm Edd}$ $>$ 10$^{-4}$, the detection rates of megamasers and disk masers are consistent with being flat functions of $\lambda_{\rm Edd}$. This may imply that, as long as  $\lambda_{\rm Edd}$ is above the threshold, the accretion efficiency may not play an important role for exciting disk maser emission (at least for X-ray selected AGN). It is likely that the maser detection rates may be different functions of $\lambda_{\rm Edd}$ for optically selected AGN because these AGN could contain a substantial population of X-ray weak, unobscured Seyfert 2 galaxies (Georgantopoulos \& Akylas 2010) at low luminosity and accretion rates (i.e. $\lambda_{\rm Edd}$ $<$ 10$^{-2}$), for which the maser detection rates are expected to be low.

\begin{figure*}
\begin{center} 
\vspace*{0 cm} 
\hspace*{-2.0 cm} 
\includegraphics[angle=0, scale=0.65]{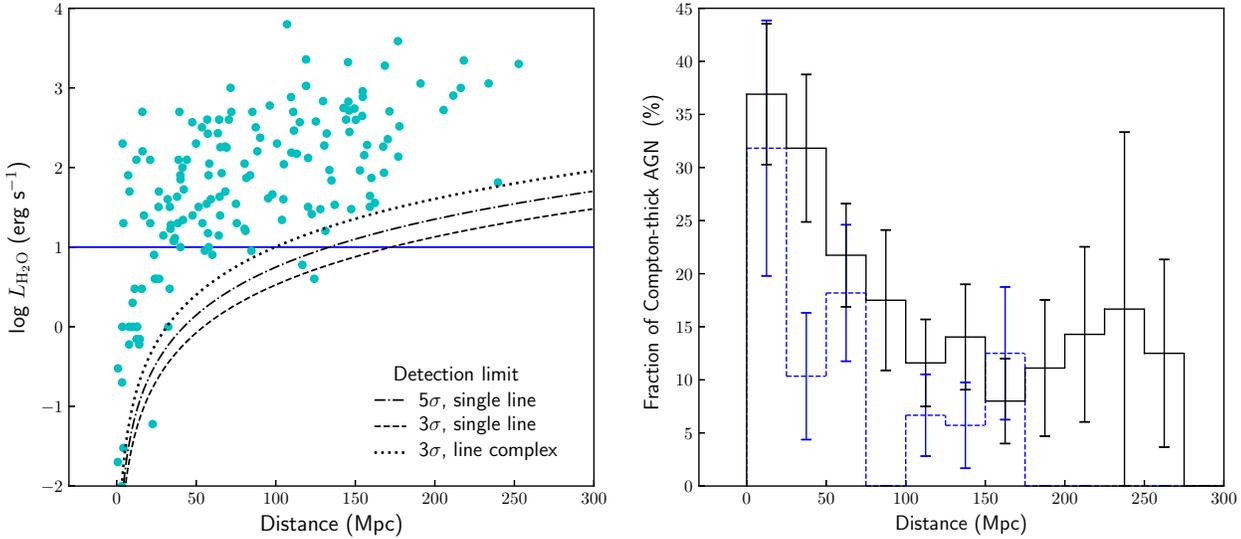}
\vspace*{0.0 cm} 
\caption{{\bf Left panel:}  The isotropic maser luminosity $L_{\rm H_{2}O}$ as a function of galaxy distance for all maser sources in the GBT sample. The dashed and dot-dashed lines show the average 3-$\sigma$ and 5-$\sigma$ detection limits of the GBT maser survey. The dotted line indicates the detection limit for a line complex consisting of three individual lines, with each line detected at the 3$\sigma$ level. Such a line complex is harder to detect because a given total maser luminosity gets diluted when spread across multiple lines. {\bf Right panel:} The fraction of Compton-thick AGN candidates in the X-ray sample as a function of distance. The black solid line shows the distribution for the candidates of compton thick AGN inferred from $L^{AGN}_{12 \micron}$$-$$L^{obs}_{2-10}$ diagram (i.e. the galaxies lying below the orange line in Figure 5). The error bars illustrate the associated Poisson uncertainties. For comparison, we also plot the fraction of Compton-thick AGN drawn from the {\it Swift}/BAT catalog which provides direct and reliable $N_{\rm H}$ measurements. The blue dashed line indicate the distribution for {\it Swift}/BAT sources which have $N_{\rm H}$ $\ge$10$^{24}$ cm$^{-2}$ from X-ray spectral fitting.  }
\label{dist-stat-fig9}
\end{center} 
\end{figure*}

\section{Discussion}

\subsection{Why do megamasers prefer mid-IR luminous and X-ray obscured AGN ? }

The analysis presented here reveals that the H$_{2}$O megamaser detection rate ($R_{\rm Mmaser}$) is a strong function the column density $N_{\rm H}$ of AGN. In addition, the detection rate also increases with 12 micron AGN luminosity $L^{AGN}_{12 \micron}$. Given the expectation that it is the dusty torus conceived in the unification paradigm of AGN (Antonucci \& Miller 1985) which blocks the X-ray photons and provides the reprocessed mid-IR emission in obscured AGN, the dependence of $R_{\rm Mmaser}$ on $N_{\rm H}$ and $L^{AGN}_{12 \micron}$ suggests that H$_{2}$O megamaser emission is closely associated with AGN circumnuclear obscuration, as previously expected (see also Kuo et al. 2018).

These results support previous findings regarding links between the properties of the obscuring material and those of the water megamasers.  
Masini et al. (2016) showed that the maser disk emission is very likely to originate from a geometrically thin disk composed of a large number of molecular clouds that connect the inner edge of the torus and the outer part of the accretion disk.  Since maser emission is beamed and long path lengths are required for strong maser amplification, the detection rate of megamaser disks is then expected to be the highest when maser disks are close to being edge-on. Assuming that the disk plane for maser emission is aligned with the equatorial plane of the circumnuclear dusty torus, one can perceive that only for an edge-on maser disk the absorbing column density $N_{\rm H}$ would become especially high because one would encounter the greatest number of obscuring clouds along the line-of-sight.  According to the standard cloud distribution of the clumpy torus model proposed by Elitzur \& Shlosman (2006), the number of torus clouds intercepted by the line-of-sight can be expressed as $N_{\rm cl}$ $\propto$ exp(-$\beta^{2}$/$\sigma^{2}$), where $\beta$ is the angular offset between the observing line-of-sight and the equatorial plane, and $\sigma$ ($\sim$45$^{\circ}$) measures the angular width of the cloud distribution, which supports a correlation between the inclination angle and the the $N_{\rm cl}$ (dropping quickly when $\beta > 0^{\circ}$), and thus $N_{\rm H}$.  
Therefore, it can be expected that the more a maser disk is away from being edge-on, the shorter the amplification path length would be, corresponding to weaker the maser emission, which will be harder to detect. As the maser disk plane has a greater departure from being edge-on, i.e., larger viewing angle $\beta$, $N_{\rm cl}$ and thus $N_{\rm H}$ would decrease. 

A similar argument could also be applied to megaser emission arising from a dusty torus (e.g. NGC 3079; Elitzur \& Shlosman 2006). In such a scenario, strong maser emission can occur when a seed photon of maser emission from a torus cloud is amplified by other clouds which lie across the line-of-sight. Based on the cloud distribution described above, the chance for such alignments would be greatest when $\beta$ is small. Therefore, the strength of the megamaser emissions from torus clouds, and the associated detection rate, will again be strongly dependent on the viewing angle of the torus, which is well correlated with the absorbing column density $N_{\rm H}$.

On the other hand, a large number of clouds along the ling-of-sight is clearly not sufficient for generating strong maser emission. This is because maser excitation can become significant only when the gas temperature $T_{\rm gas}$ is greater than $\sim$400 K, which is the temperature above which the reaction network for generating H$_{2}$O molecules becomes very efficient and can lead to a water abundance of 10$^{-4}$ relative to the density of H nuclei (Neufeld, Maloney, \& Conger 1994; Neufeld \& Maloney 1995).  
The key to raise the gas temperature above $\sim$400 K is the X-ray heating rate, which depends on both the illuminating hard X-ray flux $F_{\rm X}$ and the column density $N_{\rm H}^{cl}$ of X-ray shielding material for a maser emitting cloud. Based on the relationship between $T_{\rm gas}$, $F_{\rm X}$, and $N_{\rm H}^{cl}$ obtained by Neufeld, Maloney, \& Conger (1994), it can be inferred that for a population of disk (or torus) clouds with a given $N_{\rm H}^{cl}$ distribution, the higher the X-ray flux is, the higher the fraction of the clouds that will be heated to sufficient $T_{\rm gas}$ for maser emission will be.  Therefore, if an AGN has a higher X-ray luminosity $L_{X}$ (i.e., higher 12$\micron$ AGN luminosity), the megamaser emission is also more likely to become stronger thanks to the higher number of clouds at suitable $T_{\rm gas}$ for maser emission and amplification, leading to higher megamaser detection rates. 


\subsection{The Effect of Galaxy Distance on Maser Detection}
\subsubsection{The Role of the Sensitivity Limit}
Galaxy distance has a subtle effect on maser detection, and this effect may play different roles for optically-selected and X-selected AGN. From the right panel of Figure 1, we can see that the maser detection rates of the X-ray sample drop quickly as a function of distance $D$. In addition, the detection rates of megamasers and disk masers increase by a factor of a few relative to the GBT sample at $D{\sim}50$ Mpc. The rapidly decreasing trend in the megamaser detection rates -- which are substantially enhanced at short distances -- is in sharp contrast to the {\it flat} detection rate distribution for the GBT sample (see Figure 1 in Kuo et al. 2018), which shows that the detection rates of megamasers and disk masers are nearly constant ($\sim$2.7\% for megamasers and $\sim$0.9\% for disk masers) from $D{=}0{\sim}180$ Mpc. Given that the X-ray sample is a subset of the GBT sample, we expect that the dramatically different appearances in the distributions of the megamaser detection rates are the result of selection effects. It is likely that the X-ray selection gives rise to a significant distance effect in the search for H$_{2}$O megamasers.

When looking for the causes of this difference in the detection rate distributions, we notice that the sensitivity limit for megamaser/disk maser detection in an H$_{2}$O maser survey cannot fully account for what we see. In the left panel of Figure \ref{dist-stat-fig9}, we plot the H$_{2}$O maser luminosities of all maser sources in the GBT sample as a function of $D$. The dashed and dot-dashed lines show the 3-$\sigma$ and 5-$\sigma$ detection limits of the GBT maser survey\footnote{The RMS noise of each source in the GBT maser survey can be found at http://www.gb.nrao.edu/$\sim$jbraatz/H2O/sum\_dir\_sort.txt. The mean RMS of the survey is 4.9 mJy for a channel width of 0.33 km~s$^{-1}$ (24.4 kHz).} presented in Kuo et al. (2018). Here, we define the detection limits as the sensitivity needed for detecting a single maser line with a linewidth broader than 1 km~s$^{-1}$ (i.e. the typical width of an H$_{2}$O maser line) at 3$\sigma$ or 5$\sigma$ significance. 

From the trends in the detection limits, one can infer that sensitivity starts to substantially affect the detection rate of megamasers (i.e. $L_{\rm H_{2}O}$ $\ge$ 10 $L_{\odot}$, the sources above the horizontal blue line in the plot) only when $D$ is greater than $\simeq$170 Mpc, suggesting that one could in principle detect nearly all megamasers up to $D{\sim}170$ Mpc, given  the sensitivity level achieved in the GBT maser survey.  This would not only explain why the megamaser detection rate is nearly constant for the GBT sample within $D$ $\simeq$180 Mpc, but it would also suggest that the rapidly decreasing trend in the maser detection rates for the X-ray sample results from a different cause.

This conclusion would not change dramatically if one imposed a stricter definition for the detection limit by taking into account the fact that a maser detection usually appears in the form of line complexes. A line complex tends to be more difficult to detect for a given total maser luminosity because the total maser flux gets diluted when spread across multiple lines, resulting in each line having a smaller signal-to-noise ratio. In Figure \ref{dist-stat-fig9}, the dotted line represents the detection limit for a maser line complex consisting of at least 3 individual maser lines, with each line detected at the 3$\sigma$ level. Based on this definition, one would start to lose detections of megamasers only when $D$ is greater than 100 Mpc. The megamaser detection rate will still be a flat function of $D$ within 100 Mpc, inconsistent with the quick drop in the detection rates seen in Figure \ref{mstat}. 

Given the close relationship between megamasers/disk masers and the highly obscured/Compton-thick AGN, we argue that the reason why the detection rates of megamasers and disk masers in the X-ray sample are strongly decreasing functions of $D$ is because the fraction of high column density AGN decreases rapidly with $D$ for a sample of X-ray selected AGN. Due to strong photoelectric absorption, it is expected that the X-ray emission from highly obscured and Compton-thick AGN will get significantly suppressed (e.g. Treister, Urry, \& Virani 2009). These suppressed X-ray sources become harder to detect at greater distances in a sensitivity-limited X-ray survey, leading to a decreasing fraction of heavily obscured AGN as $D$ increases. As a result, if one searches for masers from a sample of X-ray selected sources in which  the  fraction  of  highly  obscured  AGN  becomes  smaller  because  of  the  X-ray  sensitivity limit,  the  expected  maser  detection  yields  will also be lower because the search will be conducted primarily among lower column density AGN.

Indeed, when we examine the Compton-thick candidates seen in the $L^{AGN}_{12 \micron}$$-$$L^{obs}_{2-10}$ diagram, we do see that the fraction of these Compton-thick AGN drops quickly with $D$ (e.g., right panel of Figure \ref{dist-stat-fig9}), and this trend is clearly correlated with the detection rate distributions seen in Figure \ref{mstat}. This correlation suggests that the detectability of heavily-obscured X-ray sources as a function of distance is likely to be the dominant cause for the decreasing trend of the maser detection rates for the overall samples of X-ray selected AGN.

\begin{figure*}
\begin{center} 
\vspace*{0 cm} 
\hspace*{0.0 cm} 
\includegraphics[angle=0, scale=0.7]{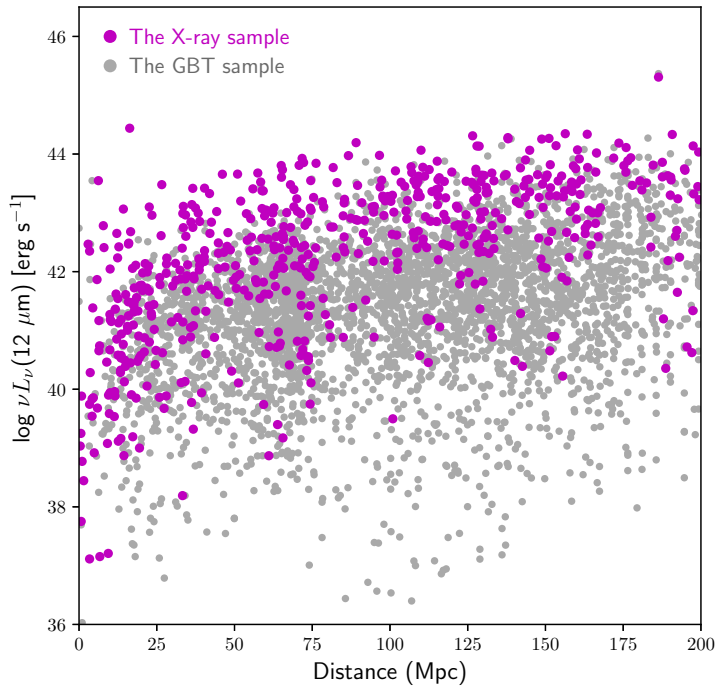}
\vspace*{0.0 cm} 
\caption{The 12 $\micron$ AGN luminosity as a function of distance. The magenta and grey dots show galaxies in the GBT-X-ray sample and the whole GBT sample of galaxies without X-ray  information, respectively.}
\label{mir-dist-fig10}
\end{center} 
\end{figure*}

\subsubsection{The Role of mid-IR AGN Luminosity}
While the decreasing fraction of Compton-thick AGN in the X-ray sample provides an explanation for the declining trend of the megamaser detection rate, it cannot easily explain the enhanced peak in the megamaser detection rate distribution at $D=$ 25$-$50 Mpc, suggesting that the X-ray selection may introduce a secondary distance effect in the search for H$_{2}$O megamasers. After examining factors that might enhance megamaser detection rates as a function of distance, we note that this peak could be associated with the dependence of the megamaser detection rate on the mid-IR AGN luminosity.

Figure \ref{mir-dist-fig10} shows the 12 $\micron$ AGN luminosities $L^{AGN}_{12 \micron}$ as a function of $D$ for the sources in the  GBT-X-ray sample (the magenta dots) and the whole GBT sample regardless of the X-ray detection (the grey dots). The mid-IR AGN luminosities for the GBT sample are obtained in exactly the same way as described in Section 2.2. As expected, the AGN luminosities for both samples increase as a function of distance. However, because of the sensitivity limits, the X-ray surveys tend to pick the most luminous fraction of the AGN in the GBT sample when the distance is greater than $D$ $\sim$25$-$50 Mpc.  
Note also that the X-ray galaxies beyond 25 Mpc are gradually dominated by sources with $L^{AGN}_{12 \micron}$ $\gtrsim$10$^{42}$ erg~s$^{-1}$, whereas the majority of the X-ray sources within 25 Mpc have $L_{12 \micron}$ $<$10$^{42}$ erg~s$^{-1}$.

As described in Section 3.1, the megamaser detection rate is an increasing function of $L^{AGN}_{12 \micron}$ when $L^{AGN}_{12 \micron}$ $<$10$^{42}$ erg~s$^{-1}$, and it increases by a factor of a few when $L_{12 \micron}$ is greater than 10$^{42}$ erg~s$^{-1}$. Therefore, one can expect that when the fraction of Compton-thick sources is nearly constant within $D$ $=$ 50 Mpc, the distance distribution of the megamaser detection rate is mainly affected by the mid-IR AGN luminosity, and it is expected to increase by a factor of a few when the distance is greater than $\sim$25 Mpc because the X-ray sources gradually have $L_{12 \micron}$ $\gtrsim$10$^{42}$ erg~s$^{-1}$. Beyond 50 Mpc, the rapidly decreasing fraction of Compton-thick sources starts to affect the detection rate distribution significantly and causes the megamaser detection rate drop. In the end, as a result of the X-ray detection limit, the X-ray selection gives rise to the quickly increasing trend of mid-IR AGN luminosity within 50 Mpc and the rapidly decreasing Compton-thick fraction beyond 50 Mpc, leading to an enhanced peak at $D{\sim}25{-}50$ Mpc in the megamaser detection rate distribution and a quick decrease when $D$ $>$ 50 Mpc.

Finally, the effect of AGN luminosity discussed here could also explain why an early maser survey conducted by Braatz et al. (1996) achieved a relatively high maser detection rate  (i.e. 11$\pm$5\%) compared with the whole GBT sample for nearby AGN located within $\sim$30 Mpc. While the majority of the galaxies in the GBT sample within 30 Mpc have $L^{AGN}_{12 \micron}$ $<$ 10$^{41}$ erg~s$^{-1}$, the majority of AGN included in Braatz et al. (1996) have $L^{obs}_{2-10}$ $>$ 10$^{41}$ erg~s$^{-1}$ (i.e. $L^{AGN}_{12 \micron}$ $>$ 2.5$\times$10$^{41}$ erg~s$^{-1}$; see Falocco et al. 2014)\footnote{The AGN sample in Braatz et al. (1996) was drawn either from the V$\acute{\rm e}$ron-Cetty \& V$\acute{\rm e}$ron catalog (1991) or from Huchra's catalog of AGN (Huchra 1993; private communication). While the range of $L^{obs}_{2-10}$ for sources in the V$\acute{\rm e}$ron catalog can be found in Falocco et al. (2014), there are no published X-ray luminosities for  Huchra's AGN sample obtained via private communication. Therefore, for the argument presented in this paragraph, we assume that an early AGN sample such as Huchra (1993) mainly consists of AGN as luminous as the sources included in the V$\acute{\rm e}$ron-Cetty \& V$\acute{\rm e}$ron catalog. }. Given that the megamaser detection rate ranges between 4-14\% when $L^{obs}_{2-10}$ is greater than 10$^{41}$ erg~s$^{-1}$ (see Figure 4), it is conceivable that an early maser survey which selected primarily luminous AGN could achieve a higher detection rate than the general GBT survey for nearby optically active sources.

\subsection{Prediction}
In Section 3, we showed that the $L^{AGN}_{12 \micron}-L^{obs}_{2-10}$ diagram provides a convenient way to select AGN that are simultaneously heavily-obscured and mid-IR luminous, for boosting the maser detection rates. 
This method can be easily applied to a new sample of X-ray sources, and it allows us to predict how many megamasers and disk masers we could discover in the near future.

To establish the new X-ray sample for making such predictions, we first collect all X-ray AGN from the largest pools of X-ray sources with spectral-fitting results available -- the {\it Swift}/BAT AGN catalog and the {\it XMM-Newton} spectral-fit database -- followed by removing all sources that have already been observed by the GBT as part of a maser search. Because of the sensitivity limit of VLBI observations for black hole mass measurements (Kuo et al. submitted), we only include galaxies within $z$ $<$ 0.07 ($\sim$300 Mpc) in our new X-ray sample of galaxies that have never been surveyed for emission in 22 GHz.  Since the {\it XMM-Newton} catalog does not provide spectroscopic redshifts, prior to selecting X-ray sources within $z$ $<$ 0.07 we cross-match the {\it XMM-Newton} catalog with 12 galaxy redshift surveys\footnote{These redshift surveys include : 1. Galaxy Zoo (Lintott et al. 2008); 2. 2MASS Redshift Survey (Huchra et al. 2012); 3. 2dF Survey (Colless et al. 2001); 4. 6dF Survey (Jones et al. 2009); 5. RC3 catalog (de Vaucouleurs et al. 1991); 6. Galaxy Zoo 2 (Willett et al. 2013); 7. CfA Redshift Catalog (ZCAT; Huchra et al. 1995); 8. CFA2S (Huchra et al. 1999); 9. Dark Energy Survey (Abbott et al. 2018); 10. Point Source Catalog Survey (PSCz; Saunders et al. 2000); 11. Updated Zwicky Catalog (UZC; Falco et al. 1999); 12. SDSS DR 14 (Abolfathi et al. 2018).}  to obtain galaxy redshifts and to maximize the number of X-ray sources with associated spectroscopic redshifts.

After assembling all X-ray sources within $z$ $=$ 0.07, we collect the broadband UV-to-mid-infrared photometry in exactly the same way as described in Section 2. In addition, we perform SED-fitting with MAGPHYS to measure the 12 $\micron$ AGN luminosity. The total number of X-ray sources having both spectroscopic redshifts and the photometric data necessary for the SED-fitting is 628, with 441 (70\%) of them lying within $z$ $=$ 0.04 ($\sim$170 Mpc). 

\begin{figure*}
\begin{center} 
\vspace*{0 cm} 
\hspace*{0.0 cm} 
\includegraphics[angle=0, scale=0.7]{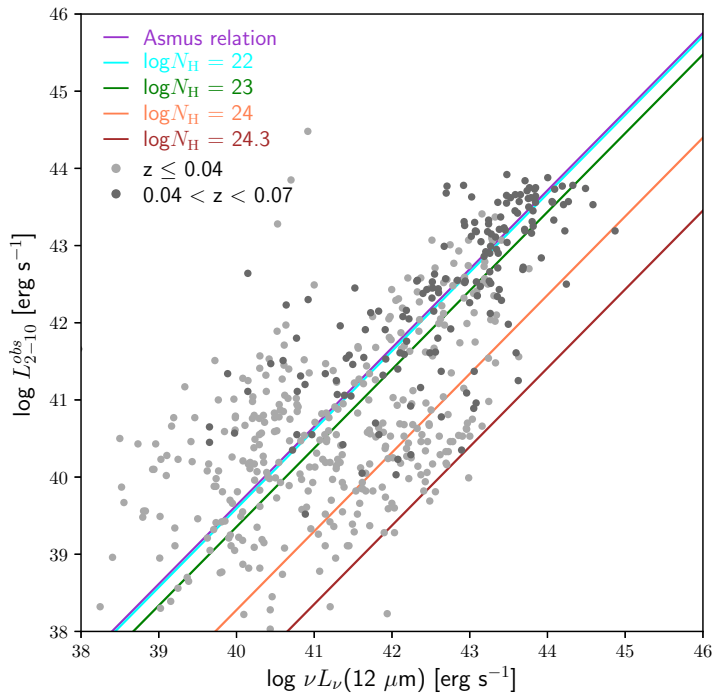}
\vspace*{0.0 cm} 
\caption{The $L_{\rm MIR}-L_{\rm X}$ diagram for X-ray sources from the {\it Swift}/BAT and the {\it XMM-Newton} catalogs which have not yet been searched for H$_{2}$O masers with the GBT. The grey and dark grey circles refer to X-ray sources with $z$ $\le$ 0.04 and 0.04 $<$ $z$ $<$ 0.07, respectively.}
\label{asmus-predict-fig11}
\end{center} 
\end{figure*}

In Figure \ref{asmus-predict-fig11}, we plot $L^{AGN}_{\rm 12 \micron}$ against $L_{2-10}^{obs}$ for the new sample of non-GBT-X-ray sources. The grey and dark grey symbols represent X-ray sources within and beyond $z$ $=$ 0.04, respectively. From this plot, we can see that the number of X-ray sources having $N_{\rm H}$ $\gtrsim$ 10$^{24}$ cm$^{-2}$ is substantially smaller for galaxies beyond $z$ $=$ 0.04, consistent with the distance effect discussed in Section 4.2.

To achieve high maser detection rates and completeness rates simultaneously for this new sample, we adopt the criterion $N_{\rm H}$ $\ge$ 10$^{23}$ cm$^{-2}$ for sample selection. The number of X-ray sources satisfying this condition is 233, which then corresponds to an expected numbers of new megamaser and disk maser detections of 35$\pm$5 and 15$\pm$3, respectively. Alternatively, we can also adopt the criterion $N_{\rm H}$ $\gtrsim$ 10$^{24}$ cm$^{-2}$ to select galaxies, and this leads to a total number of 96 sources satisfying the condition, corresponding to new detections of  19$\pm$4 megamasers and 9$\pm$3 disk masers. For the rest of the selection criteria presented in this paper, we list the predicted numbers of disk masers in Column 10 of Table 4. 

We note that in the above predictions we make the assumption that all of the megamasers and disk masers can be detected up to $z$ $=$ 0.07. For this assumption to be valid, we require that the integration time for sources between $z$ $=$ 0.04$\sim$0.07 be increased by a factor of 4 on average (i.e. $\sim$40 minutes per source) in order to detect all megamasers above the 3 $\sigma$ level. Fortunately, the number of heavily-obscured AGN between $z$ $=$ 0.04$\sim$0.07 is $\lesssim$48 ($\sim$21\% of the targets) and thus, even if one increases the integration time for these higher redshift sources, the total amount of observing time needed for the 233 X-ray AGN amounts to only $\sim$63 hours with the GBT.  This exposure time requirement is about a factor of 3 smaller than the typical GBT time awarded to the MCP per year before $\sim$2013 (200 hours), while the maser detection is at least 5 times more efficient in terms of the detection rates.

\section{Conclusion}
To reach the ultimate potential of H$_{2}$O megamasers for measuring a percent-level \ho, as well as for a significant increase in the number of accurately measured SMBH masses, it is important to enhance the efficiency with which we can discover new megamaser disk systems. While the mid-IR color cuts for selecting red/dusty AGNs presented in Kuo et al. (2018) reveal a dramatic increase (at least a factor of four) in the maser detection rates, this method is compromised by a small maser completeness rate ($\sim$30\% for the criteria leading to the highest detection rates) and a small fraction ($<$10\%) of AGN complying with the criteria. The low completeness rate results from the fact that the mid-IR emission in the majority ($\sim$60$-$70\%) of maser galaxies is dominated by the host galaxy component, making these ``\emph{WISE} blue'' galaxies easily missed by our mid-IR selection criteria.

In this paper, we demonstrate that by incorporating hard X-ray information in the analysis and performing SED-fitting, we can minimize the impact of the host galaxy emission on identifying hosts of water megamasers. Instead of selecting dusty (heavily obscured) AGN based on mid-IR colors, we use the ratio of the observed X-ray to mid-IR AGN luminosity to infer the obscuring column density $N_{\rm H}$, which can be successfully used as a proxy for the highly sought megamasing systems. This method is particularly useful for discovering heavily-obscured AGN candidates from X-ray surveys probing $\le$10 keV bands. Our analysis shows that the detection rates of megamasers and disk masers primarily depend on $N_{\rm H}$, with the majority (93\%) of them having $N_{\rm H}$ $\ge$ 10$^{23}$ cm$^{-2}$. 

Because of the obscured nature of the megamaser systems, selecting galaxies with $N_{\rm H}$ $\ge$ 10$^{23}$ cm$^{-2}$ and $N_{\rm H}$ $\ge$ 10$^{24}$ cm$^{-2}$ leads to impressively high detection rates: $\sim$15$-$28\% for megamasers and $\sim$6$-$13\% for disk masers, which are $\sim 5-10$ times higher than the typical rates achieved in the MCP survey, and the compromise in the completeness rates is mild.  With respect to the X-ray sample, the maser completeness rates are as high as $\sim$95\% and $\sim$50\% if one chooses galaxies with $N_{\rm H}$ $\ge$ 10$^{23}$ cm$^{-2}$ and $N_{\rm H}$ $\ge$ 10$^{24}$ cm$^{-2}$, respectively. Making a cut of $L^{AGN}_{12 \micron}$ $\gtrsim$10$^{42}$ erg~s$^{-1}$ would lead to only a mild decrease in the completeness rates. This shows that {\sl X-ray-plus-mid-IR selection} is not only effective for significantly boosting the maser detection rates, but also efficient at discovering the bulk of megamaser galaxies among the entire AGN population.  
 
The maser galaxy selection methods established in this paper can be easily applied to current and future X-ray AGN surveys for discovering more disk maser systems. When applied to the 628 X-ray sources available in the {\it Swift}/BAT AGN catalog and the {\it XMM-Newton} spectral-fit database within $z = 0.07$, we predict the detection of $\sim$15 new disk megamaser systems by new 22 GHz surveys. Although this number is only about 23\% of the number of new disk masers ($\sim$70) needed to achieve a 1\% Hubble constant measurement with current instrumentation (e.g., Kuo et al. 2018), it is certain that the number of disk maser candidates will continue to increase as both {\it Swift}/BAT and {\it XMM-Newton} AGN catalogs keep growing, while future X-ray observatories such as STROBE-X (Ray et al. 2018) will discover more heavily-obscured AGN. 
 
Therefore, along with the detection rate enhancement methods proposed by Kuo et al. (2018), it is promising that we will gather enough disk maser candidates to make a 1\% $H_{0}$ determination, along with the very desirable black hole masses at percent level accuracy, when the Next Generation Very Large Array (ngVLA) is well established and included in the VLBI, which is the main tool used to provide the necessary highly accurate astrometric measurements for these goals.

\acknowledgements

We gratefully thank the anonymous referee for a thorough review that improved this manuscript.  
This publication is supported by Ministry of Science and Technology, R.O.C. 
under the project 108-2112-M-110-002.   
A.C. acknowledges support from the National Science Foundation under Grant No. AST 1814594.  This work has made use of data products from the \emph {Wide-field Infrared Survey Explorer (WISE)} and the SDSS, and the 2MASS. \emph {WISE} is a joint project of the University of California, Los Angeles, and the Jet Propulsion Laboratory/California Institute of Technology, funded by the National Aeronautics and Space Administration. 
  
SDSS is managed by the Astrophysical Research Consortium for the Participating Institutions of the SDSS-III Collaboration including the University of Arizona, the Brazilian Participation Group, Brookhaven National Laboratory, Carnegie Mellon University, University of Florida, the French Participation Group, the German Participation Group, Harvard University, the Instituto de Astrofisica de Canarias, the Michigan State/Notre Dame/JINA Participation Group, Johns Hopkins University, Lawrence Berkeley National Laboratory, Max Planck Institute for Astrophysics, Max Planck Institute for Extraterrestrial Physics, New Mexico State University, New York University, Ohio State University, Pennsylvania State University, University of Portsmouth, Princeton University, the Spanish Participation Group, University of Tokyo, University of Utah, Vanderbilt University, University of Virginia, University of Washington, and Yale University. 

The Two Micron All Sky Survey (2MASS) is a joint project of the University of Massachusetts and the Infrared Processing and Analysis Center/California Institute of Technology, funded by the National Aeronautics and Space Administration and the National Science Foundation. This research has made use of the NASA/IPAC Extragalactic Database (NED) which is operated by the Jet Propulsion Laboratory, California Institute of Technology, under contract with the National Aeronautics and Space Administration. Funding for SDSS has been provided by the Alfred P. Sloan Foundation, the Participating Institutions, the National Science Foundation, and the U.S. Department of Energy Office of Science. The SDSS web site is http://www.sdss.org/.

\end{document}